\documentclass[journal]{IEEEtran}
\IEEEoverridecommandlockouts
\usepackage{graphicx}
\def\BibTeX{{\rm B\kern-.05em{\sc i\kern-.025em b}\kern-.08em
    T\kern-.1667em\lower.7ex\hbox{E}\kern-.125emX}}
\usepackage{cite}
\usepackage{mathtools}
\usepackage{blindtext}
\usepackage{amsmath,amssymb,amsfonts}
\usepackage{algorithm}
\usepackage{algpseudocode}
\usepackage{setspace}
\usepackage{subcaption}	
\usepackage{textcomp}
\usepackage{balance}
\usepackage{xcolor}
\usepackage{multirow}
\usepackage{stfloats}
\usepackage{amsmath}
\usepackage{array}
\usepackage{url}
\usepackage{soul}
\usepackage{tabularx}
\usepackage{bm}
\usepackage{pifont}%
\newcommand{\xmark}{\ding{55}}%
\usepackage{subcaption}
\newtheorem{rem}{Remark}
\begin{document}

\title{Correction of Channel Sounding Clock Drift and Antenna Rotation Effects for mmWave Angular Profile Measurements
}

\author{Fatih Erden,
        Ozgur Ozdemir,
        Wahab Khawaja, and
        Ismail Guvenc
\thanks{This work has been supported in part by NASA under the Federal Award ID number NNX17AJ94A and in part by DOCOMO Innovations, Inc. A 1-page version of this paper is accepted for presentation at the mmWave RCN Workshop, Boulder, CO, in July 2019~\cite{RCNpaper}.}%
\thanks{All the authors are with the Department of Electrical and Computer Engineering, North Carolina State University, Raleigh, NC 27606 (e-mail:\{ferden, oozdemi, wkhawaj, iguvenc\}@ncsu.edu).}
}
\maketitle

\begin{abstract}
Millimeter-wave (mmWave) bands will be used for the fifth generation communication systems to support high data rates. For the proper characterization of the mmWave propagation channel, it is essential to measure the power angular-delay profile (PADP) of the channel which includes angle-of-departure (AoD) and angle-of-arrival (AoA) of the multipath components (MPCs). In this paper, we first describe in detail our 28~GHz channel sounder where directional horn antennas are placed on rotating gimbals. Then, for this specific sounder class, we describe and address the following two problems in extracting the MPCs from the measurements: 1) For the channel measurements at large distances between the transmitter~(TX) and the receiver~(RX), it is not possible to generate the triggering signal for the TX and the RX using a single clock (SICL). This necessitates the use of separate clocks (SECLs) which introduces a random timing drift between the clocks. 2) As the positions of the antennas change during the scanning process, total distance traveled by the same MPC differs at each measurement. These two errors together cause missing some of the MPCs and detecting MPCs that do not exist in reality. We propose an algorithm to correct the clock drift and the errors in the MPC delays due to the rotation of the antennas. We compare the MPCs from the SICL measurement and the corrected SECL measurements using a Hungarian algorithm based MPC matching method. We show that the percentage of the matched MPCs increases from 28.36\% to 74.13\% after the correction process.
\end{abstract}


\section{Introduction}
\IEEEPARstart{F}{ifth} generation~(5G) cellular systems require high data rates which can be achieved using large bandwidths on the order of GHz~\cite{Shafi2018}. The frequencies below 6~GHz are highly occupied, which makes frequencies at millimeter-wave~(mmWave) bands attractive due to the vast amount of unused spectrum available for 5G networks~\cite{fcc,Sun2017}. Deploying wireless systems relies on accurate characterization of the statistical channel propagation in the deployment band~\cite{MacCartney}. Channel models for sub-6~GHz cellular systems are the result of extensive channel measurement campaigns performed with channel sounders, where the measurement data are used to characterize the wireless channel and to derive the statistical channel model~\cite{Liang2016}. The characterization of the wireless channel is different at different frequencies, so the statistical models derived for sub-6~GHz bands may not be accurate for mmWave bands. Therefore, channel sounders operating at mmWave bands are required to characterize the nature of radio propagation at these frequencies~\cite{salous&TWC:2016,khatun&globalsip:2018,ling&TAP:2018}.   

One of the main characteristics of the mmWave wireless channel is the high path loss. To account for this loss, high gain antennas are used. After scanning the environment for the strongest path, the antennas are aligned to improve the signal strength. Sub-6~GHz channel sounders mainly measure the power delay profile (PDP) of the channel. For mmWave channel sounders, angular profile of the channel, known as the power angular-delay profile (PADP), should also be measured. In PADP measurements, angle-of-departure (AoD) and angle-of-arrival (AoA) of the multipath components (MPCs) are also determined based on the PDP measurements~\cite{Lin2017}. 

In this paper, we use directional horn antennas to measure the PADP instead of more expensive phased arrays. Calibration of the phased arrays is also more difficult. Phased arrays are electrically, whereas horn antennas need to be steered mechanically to measure the PDPs in all directions. Mechanical steering of the directional antennas may take on the order of seconds per PDP measurement. Therefore, timing synchronization, which is required to align PDPs, is a major problem. Timing synchronization is performed by triggering the transmitter~(TX) and the receiver~(RX) simultaneously. If the triggering signal for the TX and the RX is generated by a single clock, then we can assume PDPs will be aligned in the delay domain. However, it is not possible to use a single clock when the TX and the RX need to be separated physically. In this case, the triggering signal at the TX and the RX are generated by two different clocks, and the timing drift between the two clocks cause misalignment of the PDPs. For this case, the training of one of the clocks by the other clock is required.

Besides the drifting problem, rotated directional antenna-based channel sounders suffer from another problem caused by the antenna rotations while extracting the MPCs. The channel sounders operating at mmWave bands need to have a large bandwidth. By having larger bandwidth, more MPCs can be resolved in the delay domain. At some point as the bandwidth of the sounder gets larger, the delay resolution becomes comparable to the length of the antenna. In that case, due to the mechanical rotation of the horn antenna, the MPC, arriving at a particular delay bin, can arrive at a different delay bin after the rotation of the antenna. This may cause ghost MPCs to be extracted from measured PDPs, where a ghost MPC is one that does not exist in reality, but it is extracted from the measurements by mistake.    

The rest of the paper is organized as follows. In Section~\ref{sec:lit}, we give a high-level review of the channel sounder types. In Section~\ref{sec:Sounder}, we describe our specific channel sounder setup used to perform PDP measurements. We then explain in Section~\ref{sec:AngularSounder} how we modify this channel sounder setup in order to perform PDP measurements at different TX and RX angles. In Section~\ref{sec:Measurements}, we describe how we extract MPCs from these measurements to obtain the PADP of the channel. Section~\ref{sec:Correction} discusses the proposed correction algorithms due to clock drift and antenna rotation effects as well as Hungarian matching algorithm to match extracted MPCs from single-clock and separate-clock measurements. Measurement results are presented in Section~\ref{Sec:Results}, and the paper is concluded in Section~\ref{sec:Conc}.\looseness=-1

\section{Literature Review}
\label{sec:lit}
Channel sounders that perform PADP measurements can be classified into three categories: 1) sounders that use a horn antenna on a rotating gimbal~\cite{MacCartney,Lin2017}, 2) sounders that use multiple horn antennas connected to a multiplexing switch~\cite{Papazian2016}, and 3) sounders that use phased array antenna~\cite{Bas2018}. Gimbal-based sounders are simple and low cost, but the measurements take longer due to the physical rotation of the antennas. In order to perform measurements faster, multiple horn antennas that span the entire space are switched by a fast multiplexer. Additional hardware increases the cost and the space the sounder takes. Sounders with phased array antennas can perform fast measurements. However, they are also expensive, and it is more challenging to perform calibrated power measurements as each beam can have different gain. Table~\ref{Tab:sounders} summarizes the pros and cons of different sounder types.\looseness=-1

\begin{table}[t]
\footnotesize
\renewcommand{\arraystretch}{1.2}
\caption {Pros/Cons of Different Channel Sounder Types. }
\label{Tab:sounders}
\resizebox{0.49\textwidth}{!}
{\begin{tabular}{c|c|c}
\hline
 Sounder type & Pros &  Cons \\\hline
Gimbal~\cite{MacCartney,Lin2017} & simple, low cost & slow  \\
Switching antenna~\cite{Papazian2016} & fast & expensive, bulky  \\
Phased array~\cite{Bas2018} & fast &  expensive, needs calibration \\\hline
\end{tabular}}
\end{table}

There have been several studies in the literature that highlight the clock drift problem. In some cases, the duration of the measurement is short, and the clock drift may not adversely affect the measurements. In~\cite{Lin2017}, a directional horn antenna with a rotating gimbal is used only at the RX side. This limits the measurement duration to 15~minutes and, the correction of the clock drift is not performed. However, in the case when both the TX and the RX need to sweep the environment, the measurement duration will get longer, and the setup will suffer from clock drift. In~\cite{Bas2018}, phased array antennas are used where the total duration of the measurement is on the order of milliseconds, and hence the effect of clock drift can be ignored. However, phased arrays are more expensive and difficult to calibrate. In~\cite{Novotny2016}, Rubidium (Rb) clocks are used to improve the clock drift during measurements in a manufacturing environment, but the drift due to Rb clocks is not further corrected. In~\cite{Papazian2016}, an omnidirectional antenna is used at the TX, and 16 directional horn antennas are used at the RX where a multiplexer switches the antenna for AoA measurements. This setup can only measure the AoA of the MPCs however, the clock drift does not affect measurements due to fast switching rate of the antennas. \looseness=-1

In~\cite{MacCartney}, the clock drift is significant, so it needs to be corrected in post-processing. It is reported that, during training, line-of-sight (LOS) MPC of the PDP drifts less than 1~ns over a time duration of 6 hours when Rb clocks are used. However, when the training cable is disconnected, the drift is reported to be 7 ns after 2 hours of operation, and it increases non-linearly thereafter exceeding 40~ns after 5 hours. For drift tracking, it is indicated that reference PDPs at a fixed TX-RX antenna pointing angle are recorded without detailed explanation of drift correction in the post-processing. 

\begin{figure*}[t!]
\centering
\centerline{\includegraphics[trim=.5cm 9cm 8.5cm 0.35cm, clip,width=0.9\textwidth]{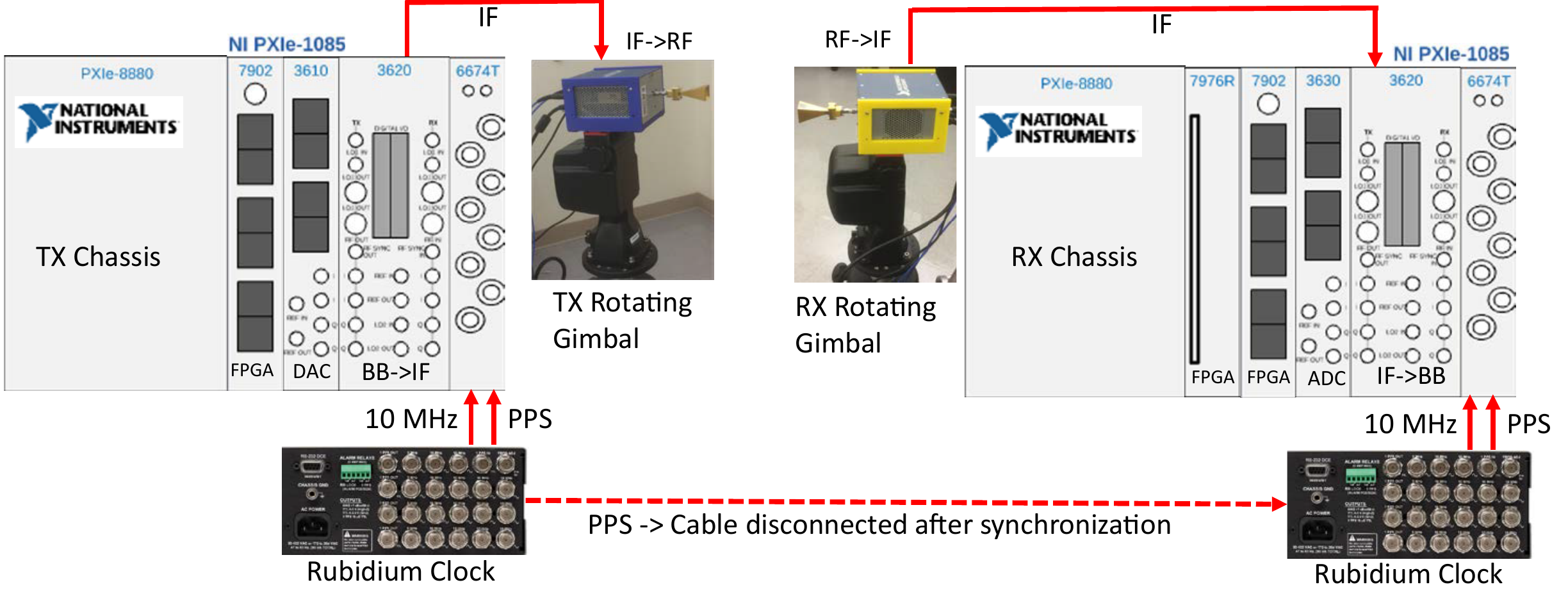}}
\caption{28~GHz channel sounder TX and RX hardware setup with two Rubidium clocks. The clocks will lose synchronization for long measurement periods.}\label{Fig:pxisetup}
\end{figure*}

Based on the above discussion, the contributions of this paper are as follows: 
\begin{enumerate}
    \item We describe in detail how the PADP can be obtained from the PDP measurements performed at different angles by rotating the directional TX and RX antennas. This includes how we modify the channel sounder operation in terms of hardware and software to measure the angular profile.
    \item We model and formulate the problems associated with the clock drift and the antenna rotations. Then, we propose an algorithm to correct the errors in delays of the peaks obtained from the PDP measurements.
Once the delays are corrected, we show how the MPCs are extracted which contain the angular profile of the channel.
\item To assess the performance improvement gained by the proposed algorithm, we perform two measurements in the same environment; one with a single clock and one with separate clocks. Single-clock measurements are disturbed only by antenna rotation effects whereas separate-clock measurements are disturbed by the clock drift as well as the antenna rotation effects. Hungarian matching algorithm is used to match the extracted MPCs from the two measurements to assess the performance improvements provided by the proposed correction algorithms.
\end{enumerate}


\section{Channel Sounder Setup}
\label{sec:Sounder}
The channel sounder hardware is based on National Instruments~(NI) mmWave system at $28$~GHz~\cite{NImmwave} and shown in Fig.~\ref{Fig:pxisetup}. It consists of NI PXIe-1085 TX/RX chassis, 28~GHz TX/RX mmWave radio heads from NI, and FS725~Rubidium~(Rb) clocks~\cite{SRS}. This sounder has been used in our previous measurements \cite{wahab_indoor,wahab_outdoor} as well. However, in~\cite{wahab_indoor}, we did not use gimbals and performed only path loss measurements, while in~\cite{ wahab_outdoor} gimbals were used, but only the AoA of the path were measured, and TX position was fixed. In this section, we describe in detail our NI channel sounder hardware and software, and in Section~\ref{sec:AngularSounder}, we introduce our modified sounder (with FLIR~PTU-D48E gimbals~\cite{FlirSystems}) that is used to perform the PADP measurements.    

\subsection{Rubidium Clocks and Drifting Problem}
As illustrated in Fig.~\ref{Fig:pxisetup}, a critical component of the channel sounder are the clocks at the TX and the RX. In particular, the 10~MHz and pulse per second~(PPS) signals generated by the Rb clocks are connected to PXIe~6674T modules at the TX and the RX. The 10~MHz signal is used to generate the required local oscillator (LO) signals, and the transmission at TX side and reception at the RX side are triggered by the PPS signal. For the proper operation of the sounder, it is essential that both the 10~MHz and the PPS signals are synchronized with each other. This synchronization is performed by a training operation: the PPS output from one of the clocks is connected to the PPS input of the other clock, and synchronization is handled by the internal clock circuitry. 

For short distance measurements it may be possible to use a single clock and obtain accurate results\footnote{In particular, we are able to perform measurements using a single clock for distances up to 30 m.}. For long-distance measurements, training needs to be performed before carrying out the measurements. Once the clocks are synchronized, the connection can be removed so that the TX and the RX can be separated from each other without any cable connecting them. In this case, the measurements are not as accurate as of the single-clock case which is mainly due to the drift of the PPS signals relative to each other. This drift is referred to as the \textit{clock drift}. The amount of drift depends on the duration of the training: the longer the training duration, the lower the drift will be. \looseness=-1

The clock drift causes the channel impulse response (CIR) to shift in the delay domain. As the channel sounder measurements to obtain the angular profile of the channel may last several hours, a significant amount of clock drift can occur during this time.  Therefore, this drift needs to be compensated to align measurements at different angles in the delay domain. This is a problem for gimbal-based sounders, but it may not be critical for switching antenna and phased array-based sounders where the total measurement time is short.

\subsection{NI Channel Sounder}
The sounder code, which is based on LabVIEW, periodically transmits a Zadoff-Chu (ZC) sequence of length 2048 to sound the channel. The ZC sequence is over-sampled by 2 when it operates on 2 GHz bandwidth, which is the case for this study, or by 4 when it operates on 1 GHz bandwidth. It is then filtered by the root-raised-cosine (RRC) filter, and the generated samples are uploaded to PXIe-7902 FPGA, as shown in Fig.~\ref{Fig:pxisetup}. These samples are sent to PXIe-3610 digital-to-analog (DAC) converter with a sampling rate of $f_s=3.072$~GS/s. The PXIe-3620 module up-converts the base-band signal to IF, and the 28~GHz mmWave radio head up-converts the IF signal to RF. 
Directional horn antennas~\cite{sageM} are connected to the mmWave radio heads at the TX and the RX sides with 17~dBi gains, and $26^{\circ}$ and $24^{\circ}$ beam-widths in the elevation and azimuth planes, respectively. 

At the RX side, 28~GHz mmWave radio head down-converts the RF signal to IF. The IF signal is down-converted to base-band at the PXIe-3620 module. The PXIe-3630 analog-to-digital converter (ADC) module samples the base-band analog signal with the sampling rate of $f_s=3.072$~GS/s. The correlation and averaging operations are performed in PXIe-7976R FPGA operation, and the complex CIR samples are sent to the PXIe-8880 host PC for further processing and saving to local disk. \looseness=-1

In the 2~GHz mode, the channel sounder provides $2/f_s\approx0.651$~ns resolution in the delay domain, and the CIR contains 2048 samples. Therefore, the maximum delay spread is $0.651$~ns $\times$  $2048=1.33$~$\mu$s. The dynamic range of the ADC is $60$~dB, and the path loss can be measured up to $185$~dB. 

\subsection{Calibration for Non-Ideal Hardware Response}
\begin{figure}[!t]
	\begin{subfigure}{0.97\columnwidth}
	\centering
	\includegraphics[width=\columnwidth]{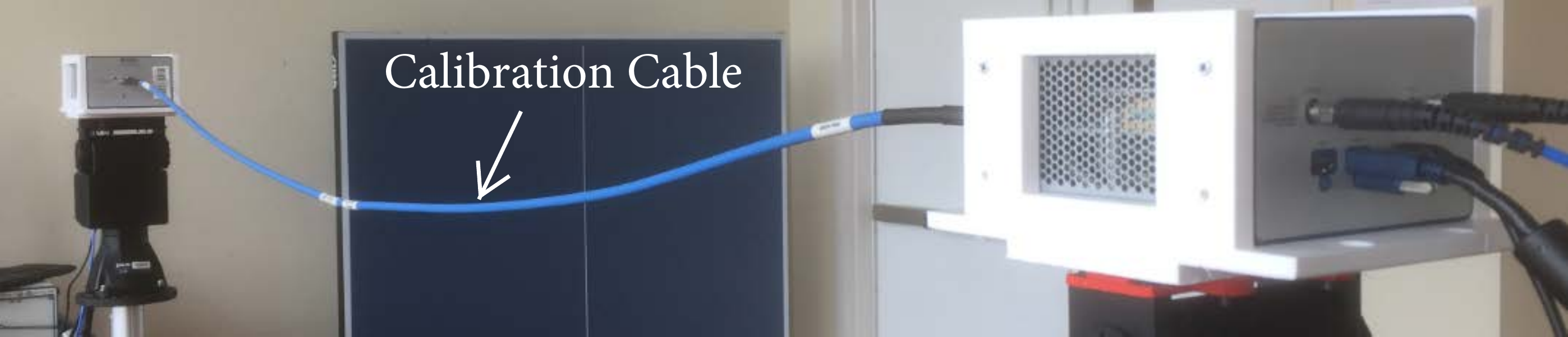} 
	\caption{} 
    \end{subfigure}			
	~\begin{subfigure}{\columnwidth}
	\centering
    \includegraphics[width=\columnwidth]{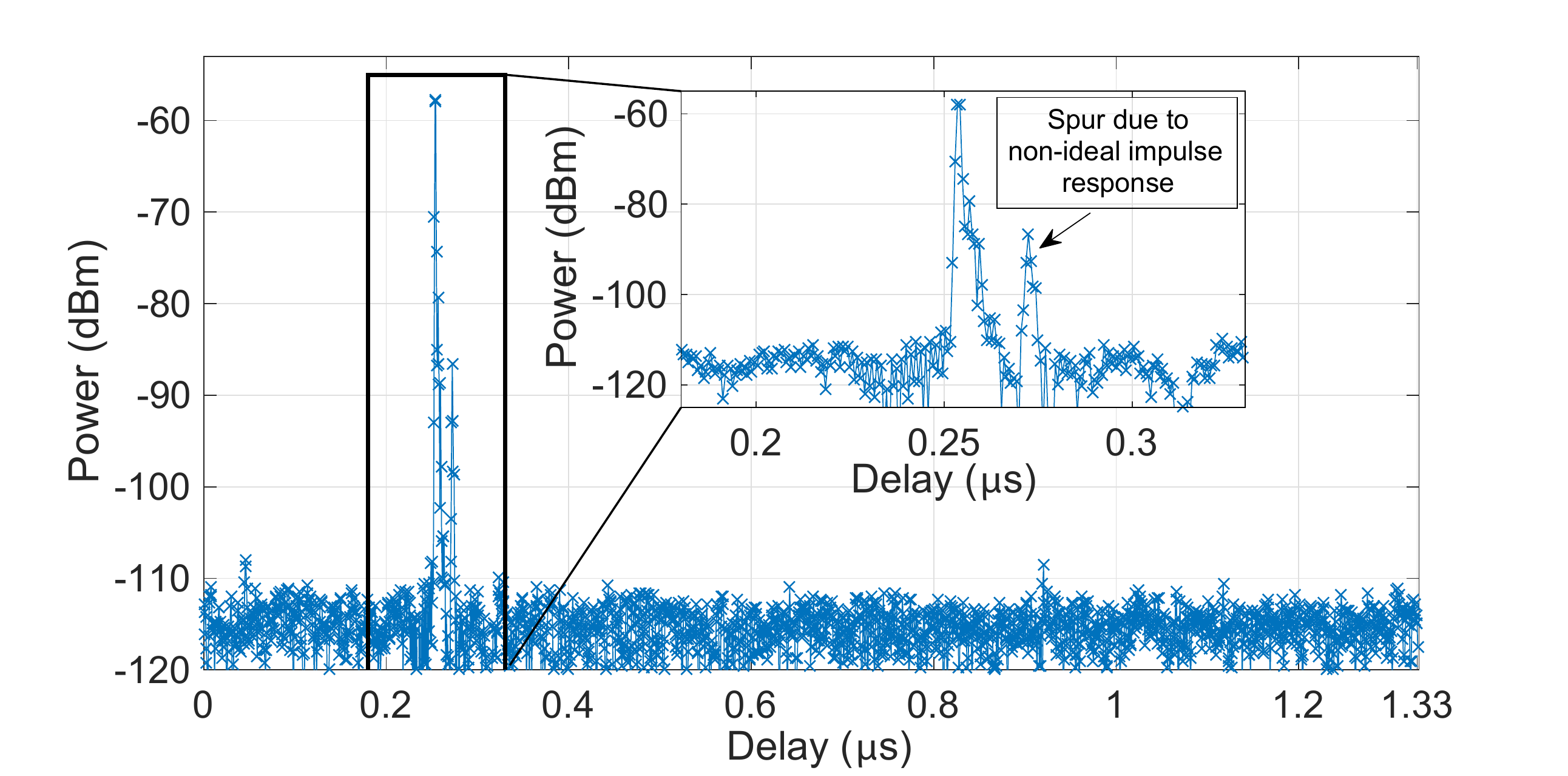}
	 \caption{}
     \end{subfigure}
	\begin{center}
     ~\begin{subfigure}{\columnwidth}
    \includegraphics[width=\columnwidth]{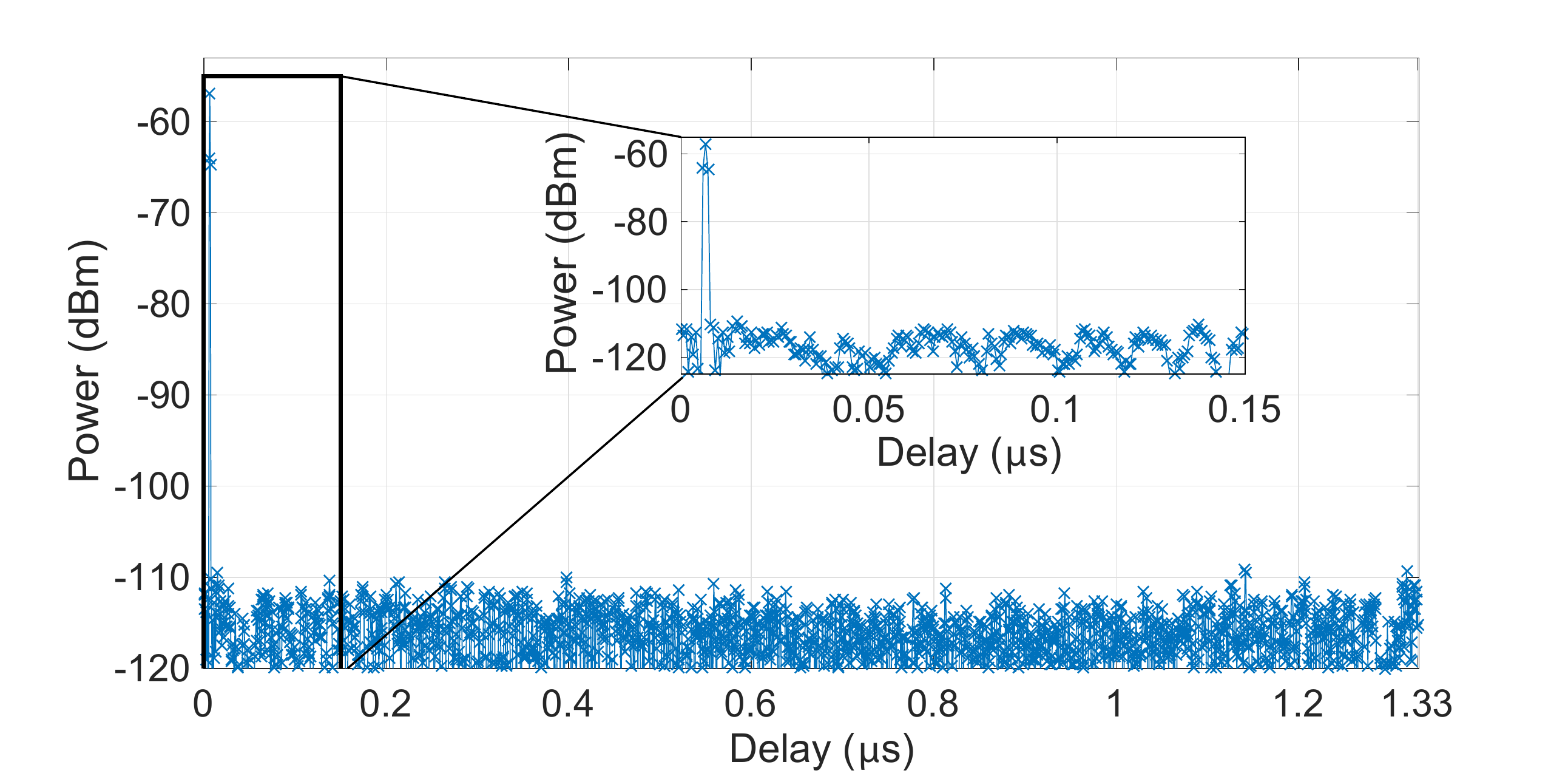}
    \caption{}
     \end{subfigure}\end{center}
     \caption{Calibration for the non-ideal hardware response: (a) calibration cable connected between TX and RX mmWave radio heads, (b) PDP obtained due to non-ideal hardware, and (c) PDP after calibration of the non-ideal hardware response. }\label{Fig:calibration}
     \vspace{-2mm}
\end{figure}

One of the challenges when performing wide-band channel sounding is that, due to non-idealities, the measurement hardware itself may introduce channel distortions which should be calibrated. These non-idealities can be measured by connecting a calibration cable between the TX/RX mmWave radio heads, as shown in Fig.~\ref{Fig:calibration}(a). The calibration cable needs to have a nearly flat response, and attenuators should be used to protect the RX. Assuming the cable and the attenuators have a flat response, the channel measured with the calibration cable attached is the non-ideal response of the measurement hardware. This response is shown in Fig.~\ref{Fig:calibration}(b) where spurs are observed due to the non-ideality of the hardware. However, as we now have the measure of this non-ideality, an equalizer can be designed to compensate for the hardware response. Fig.~\ref{Fig:calibration}(c) shows the response after the equalizer is applied where spurs are now eliminated. Applying the equalizer when the calibration cable is disconnected gives the response of the actual channel. 



\begin{figure}[t]
\centering
\includegraphics[width=0.48\textwidth]{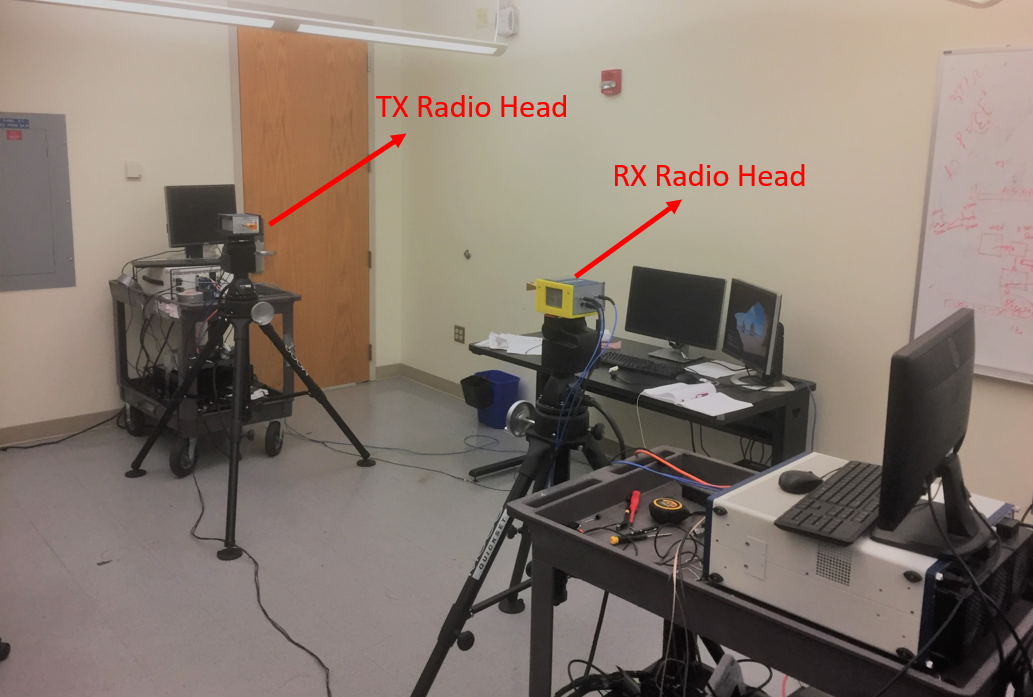}
\caption{Measurement setup to obtain the PADP of the channel. The TX and the RX mmWave radio heads are rotated using gimbals.}
\label{setup:fig}
\end{figure}

\begin{figure}[t]
\centering
\centerline{\includegraphics[scale = 0.7]{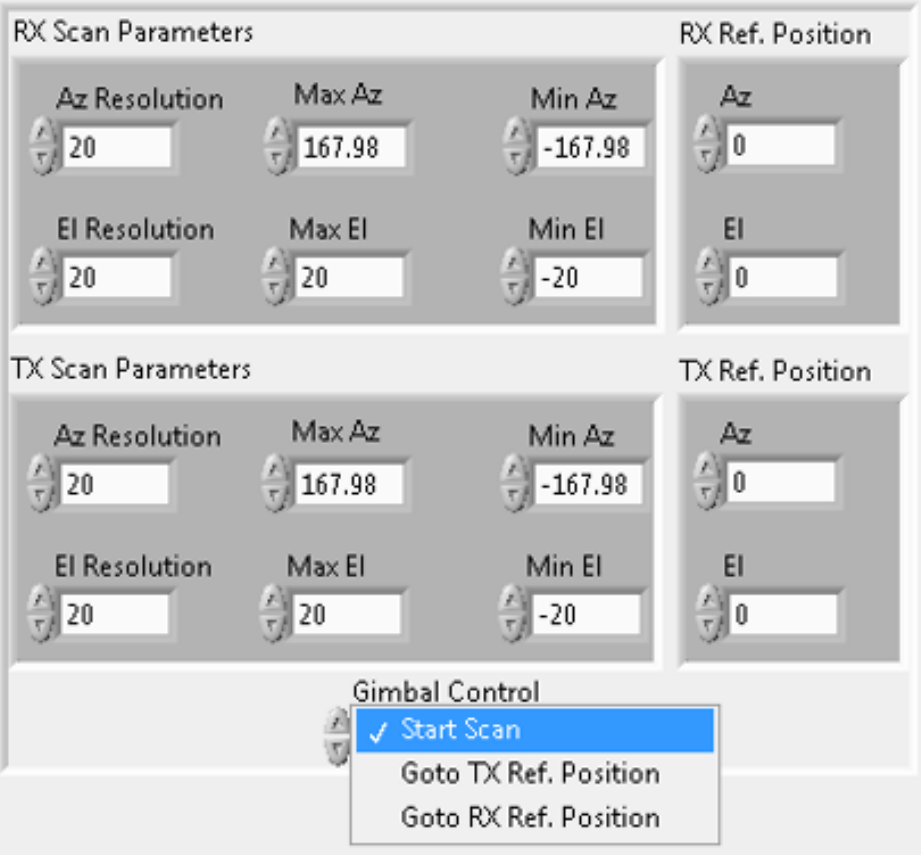}}
\caption{User interface to control the rotating gimbals.}\label{Fig:ang_param}
\end{figure}

\section{Modification to NI Channel Sounder to Measure the Angular Profile}
\label{sec:AngularSounder}
The LabVIEW based sounder code provided by the NI for the sounder in Fig.~\ref{Fig:pxisetup} can measure delays and powers of the PDP peaks measured at any position of the TX/RX antennas. We have modified this code so that we can also measure the PADP of the channel by placing the TX/RX mmWave radio heads containing directional horn antennas on rotating gimbals, as shown in Fig.~\ref{setup:fig}. Our interface, which is shown in Fig.~\ref{Fig:ang_param}, is integrated into the modified sounder code. This interface is used to control the maximum and minimum angles for azimuth and elevation angles of the gimbal at the TX/RX sides as well as the angle resolutions in the corresponding planes. The minimum and maximum scan angles of the gimbal are factory limited to $-167.98^\circ$ to $167.98^\circ$ for the azimuth plane and $-90^\circ$ to $30^\circ$ for the elevation plane.

The modified channel sounder is operated as follows:
\begin{itemize}
    \item Step 1: Move gimbal at the TX side to the next desired position.
    \item Step 2: Move gimbal at the RX side to the next desired position.
    \item Step 3: Send a trigger message from RX to TX to perform measurement at the next PPS.
    \item Step 4: Wait for the next CIR measurement.
    \item Step 5: If all the angles are visited at the RX side, go to Step 1. Otherwise, go to Step 2.  
\end{itemize}
The measurements start at a fixed reference position denoted by \textit{RX Ref. Position} and \textit{TX Ref. Position} in Fig.~\ref{Fig:ang_param}. After the first reference measurement, the scanning process begins. By default, after every 300 measurements, the reference position is revisited, then the scanning process resumes where it had left off. Once the scanning process is completed, the reference position is visited again for the last time. The reference measurements are used to track the clock drift as it is explained in Section~\ref{sec:Measurements}.

In the beginning, the TX and the RX antennas move to the minimum azimuth and elevation angles. Then the RX antenna moves from the minimum azimuth angle to the maximum azimuth angle. Next, the elevation angle of the RX antenna is increased by one in the list, and the RX antenna moves from the maximum azimuth angle to the minimum azimuth angle. The RX antenna visits all possible azimuth and elevation angle combinations, and then the TX antenna moves to the next position in its list. The measurement continues in this manner until all combinations of azimuth/elevation angles are visited by TX/RX antennas. 

The parameters shown in Fig.~\ref{Fig:ang_param}, which are used for all measurements in this paper, correspond to 3 different elevation angles $\{-20^\circ, 0^\circ, 20^\circ\}  $ and 19 different azimuth angles $\{-167.98^\circ, -160^\circ, -140^\circ, \dots, 0^\circ, \dots, 160^\circ, 167.98^\circ\}$. The total number of channel measurements $M$ is, therefore, $3 \times 19 \times 3 \times 19=3249$. Let $K$ denote the total number of reference measurements. Since, we perform reference measurements at the beginning and the end of the measurement process as well as after every 300 channel measurements, $K=1+1+10=12$. Consequently, the total number of all measurements is $K+M=3261$.

The sounder code running at the transmitter side is controlled remotely by the receiver side. In Step 1 of our modified sounder code, the receiver sends UDP messages to the transmitter to move the gimbal to the next desired position. The UDP messages are also used to send the trigger message in Step 3.\footnote{For outdoor measurements, a dedicated Wi-Fi access point device is used so that the TX/RX host PCs can obtain local IPs and communicate with each other. This setup does not require any Internet connection.} 

To measure the channel, first, we need to make sure the TX/RX antennas have moved to their next desired position. Note that even if the gimbals can move from one position to the next in less than a second, the measurement will not be performed until the PPS is ready. As a result, currently, this procedure allows us to operate as fast as one measurement per second. 

\section{Obtaining Angular Profile from Channel Sounder Measurements}
\label{sec:Measurements}

The measurements performed at a fixed position of the TX and the RX antennas provide the PDP of the channel which does not reveal the angular profile. To obtain the PADP of the channel where the AoAs and the AoDs of the MPCs over all possible TX/RX angles are estimated as well, the TX and the RX antennas scan all possible angles at azimuth and elevation planes to detect the MPCs departing from or arriving at those angles as explained in Section~\ref{sec:AngularSounder}. Then, the PADP can be expressed as: 
\begin{align}
\label{PADP:eq}
PADP(\tau, \bm{\theta}^{\mathrm{AoD}} , \bm{\theta}^{\mathrm{AoA}}) = &\sum_{n=1}^{N}  \alpha_n \delta (\bm{\theta}^{\mathrm{AoD}}-\bm{\theta}^{\mathrm{AoD}}_n)  \\ \nonumber &\times  \delta (\bm{\theta}^{\mathrm{AoA}}-\bm{\theta}^{\mathrm{AoA}}_n) \delta(\tau-\tau_{n}),
\end{align}
where $\bm{\theta}^{\mathrm{AoD}}_n = [\theta^{\mathrm{AoD,Az}}_n \,\, \theta^{\mathrm{AoD,El}}_n]^{\rm{T}}$ is the two-dimensional AoD of the $n$th MPC at the TX in the azimuth and elevation planes, $\bm{\theta}^{\mathrm{AoA}}_n = [\theta^{\mathrm{AoA,Az}}_n \,\, \theta^{\mathrm{AoA,El}}_n]^{\rm{T}}$ is the two-dimensional AoA of the same MPC at the RX, $\alpha_n$ is the path gain, $\tau_n$ is the delay of the $n$th MPC, and $N$ is the total number of MPCs considering all possible TX/RX directions.

We want to obtain the PADP of the channel by performing measurements at different positions of the TX/RX antennas. 
For the $m$th position of the TX/RX antennas, the PDP is obtained by squaring the following CIR: 
\begin{align}
\label{CIR:eq}
h_m (\tau) = \sum_{n=1}^{N} & \sqrt{\alpha_n  G_{\mathrm{TX}}(\bm{\theta}^{\mathrm{TX}}_m-\bm{\theta}^{\mathrm{AoD}}_n) G_{\mathrm{RX}}(\bm{\theta}^{\mathrm{RX}}_m-\bm{\theta}^{\mathrm{AoA}}_n)} \\ \nonumber &\times  \mathrm{e}^{j \phi_n} \delta(\tau-\tau_{n,m}-d_m)+w_m(\tau),
\end{align}
where the TX/RX antenna gains, which are functions of the antenna angles, are also considered as part of the CIR.
Here, $\bm{\theta}^{\mathrm{TX}}_m = [\theta^{\mathrm{TX,Az}}_m \,\, \theta^{\mathrm{TX,El}}_m ]^{\rm{T}}$ and  $\bm{\theta}^{\mathrm{RX}}_m = [\theta^{\mathrm{RX,Az}}_m \,\, \theta^{\mathrm{RX,El}}_m ]^{\rm{T}}$ are the angles of the TX and the RX antennas, respectively, $G_{\mathrm{TX}}(\cdot)$ and $G_{\mathrm{RX}}(\cdot)$ are the TX and the RX antenna gains, $\phi_n$ is the phase of the $n$th MPC, and $w_m(\tau)$ is the noise in the $m$th measurement. The term $\tau_{n,m}$ denotes the flight time from the TX antenna to the RX antenna of the $n$th MPC in the $m$th measurement, and $d_m$ is the cumulative sum of the random clock drift between the TX and the RX clocks up to measurement $m$.

\subsection{Clock Drift and the Delay Error Due to Antenna Rotation}
The drift term in~(\ref{CIR:eq}), $d_m$, is equal to zero when the measurements are performed using a single clock, and it increases randomly when the TX and the RX clocks are separated. To track and compensate for the clock drift, reference measurements are performed as described in Section~\ref{sec:AngularSounder}. Positions of the TX and the RX antennas for the reference measurements are selected in such a way that the LOS path is received with strong power where the distance between the TX and the RX is known as well. For non-line-of-sight (NLOS) scenarios, antenna positions for the reference measurements can be selected so that a distinctly strong MPC exists with a known TX-RX distance. \looseness=-1

\begin{figure}[t]
\centering
\centerline{\includegraphics[trim=-0.4cm 2.3cm 8.5cm 0cm, clip,width=0.49\textwidth]{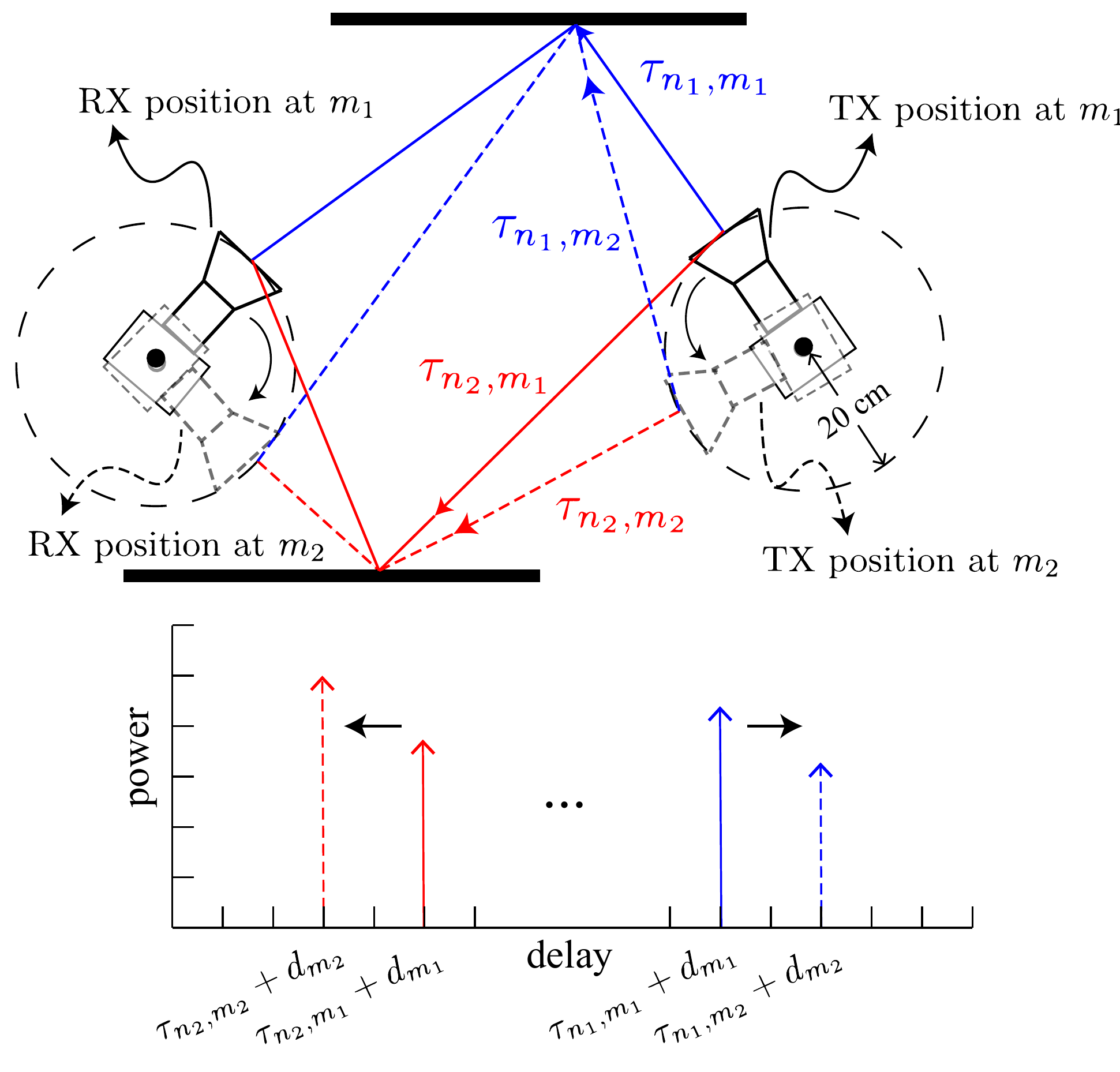}}
\caption{The effect of the clock drift and antenna rotations on the MPC parameters. Two different MPCs are illustrated at two different antenna orientations.}\label{Fig:syst}
\end{figure}

The measurements start with a reference measurement. Then, antennas start scanning the azimuth/elevation planes for the actual channel measurements. The reference measurement is repeated at certain measurement intervals (see Section~\ref{sec:AngularSounder}). After all the azimuth/elevation angles are scanned, the measurements end once again at the reference TX and RX antenna positions. The delays corresponding to a particular path at different reference measurements are used to find the clock drift at the reference measurements. These delays allow to estimate the clock drift at the channel measurements and then compensate it in the post-processing. 

A second problem is the delay error caused by the antenna rotation. The term $\tau_{n,m}$ in~(\ref{CIR:eq}) is calculated as the total distance traveled by the $n$th MPC during the $m$th measurement divided by the speed of light. As the antennas rotate, their locations will change slightly, and the distance traveled by the same MPC will vary for each measurement; therefore, the flight time is a function of the measurement index $m$ as well. 

For the measurement $m$ and MPC $n$, if $\bm{\theta}^{\mathrm{TX}}_m=\bm{\theta}^{\mathrm{AoD}}_n$ and $\bm{\theta}^{\mathrm{RX}}_m=\bm{\theta}^{\mathrm{AoA}}_n$, $\tau_{n,m}$ will have the smallest value as the antennas are aligned with the $n$th MPC both at the TX and the RX sides. Furthermore the TX/RX antenna gains for the $n$th MPC will be highest for this measurement. Denoting the minimum value of $\tau_{n,m}$ for the above scenario as $\tau_{n,\mathrm{min}}$, we define
\begin{equation}
\label{eq:delta_tau}
\Delta\tau_{n,m}=\tau_{n,m}-\tau_{n,\mathrm{min}}
\end{equation}
as the error in delay due to rotation of the antennas for the $n$th MPC at measurement $m$. 

The effect of the clock drift and the delay error caused by the rotation of the antennas is illustrated in Fig.~\ref{Fig:syst}. When the TX and the RX antennas are at the positions defined by the measurement number $m_1$, $\bm{\theta}^{\mathrm{AoD}}_{n_1}$ and $\bm{\theta}^{\mathrm{AoA}}_{n_1}$ are aligned with $\bm{\theta}^{\mathrm{TX}}_{m_1}$ and $\bm{\theta}^{\mathrm{RX}}_{m_1}$. That is, $\tau_{{n_1},\mathrm{min}}=\tau_{{n_1},{m_1}}$. Therefore, the path gain is larger and the delay is smaller for the MPC $n_1$ at measurement $m_1$ when compared with those obtained at the antenna positions at measurement $m_2$. The error in delay of MPC $n_1$ due to the change in antenna positions at measurement $m_2$ is $\Delta\tau_{{n_1},{m_2}}=\tau_{{n_1},{m_2}}-\tau_{{n_1},{m_1}}$. On the other hand, for MPC $n_2$, $\bm{\theta}^{\mathrm{AoD}}_{n_2}$ and
$\bm{\theta}^{\mathrm{AoA}}_{n_2}$ become aligned with $\bm{\theta}^{\mathrm{TX}}_{m_2}$ and $\bm{\theta}^{\mathrm{RX}}_{m_2}$ when the TX and the RX antennas move to their positions at measurement $m_2$. Thus, MPC $n_2$ appears at an earlier delay with a larger path gain at measurement $m_2$ when compared with that at the measurement $m_1$. The error in delay of MPC $n_2$ due to the antenna positions at measurement $m_1$ is $\Delta\tau_{{n_2},{m_1}}=\tau_{{n_2},{m_1}}-\tau_{{n_2},{m_2}}$, where $\tau_{{n_2},{m_2}}=\tau_{{n_2},\mathrm{min}}$. We note that the MPCs are also shifted in the delay domain by the amount of the corresponding clock drift $d_{m_1}$ or $d_{m_2}$. Here, to be able to show the effect of the antenna rotations in measuring the MPC delays, we assume that the measurement $m_2$ is performed shortly after the measurement $m_1$, i.e., $d_{m_1}\approx d_{m_2}$. We also note that the distance traveled by the MPC $n_1$ is larger than that of the MPC $n_2$ and hence $\tau_{{n_2},{m_2}}>\tau_{{n_1},{m_1}}$. We will explain the procedure for the correction of the errors due to the rotation of the antennas and the clock drift in detail in Section~\ref{sec:Correction}.

\subsection{Extracting Peaks of the PDPs}
\begin{figure}[t]
\centering
\centerline{\includegraphics[trim=0cm 0cm 0.54cm 0.5cm, clip,width=0.5\textwidth]{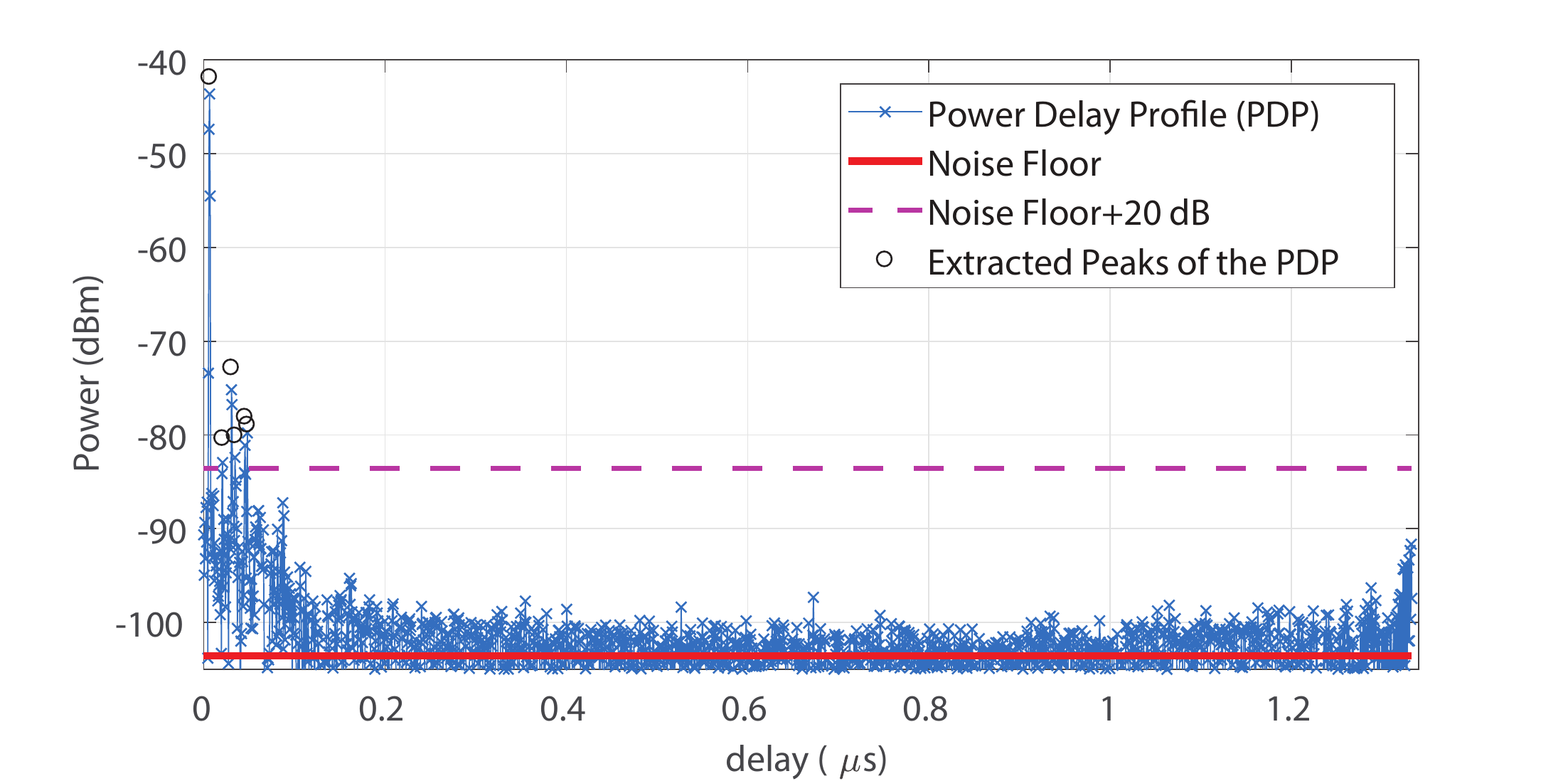}}
\caption{Example PDP for a fixed (LOS) TX/RX antenna orientation.}\label{Fig:exPDP}
\end{figure}

The peaks of the PDP for a given position of the TX/RX antennas may potentially be extracted as MPCs as described in the next subsection. An example PDP obtained at a particular measurement when the TX and the RX antennas are positioned facing each other (as in Fig.~\ref{setup:fig}) is shown in Fig.~\ref{Fig:exPDP}. To distinguish the actual PDP peak from noise, first, the noise floor is calculated as the average power of the PDP samples that are at a certain distance from the strongest MPC. Then, the peaks that are 20 dB above the noise floor are extracted. The peaks for the example PDP in Fig.~\ref{Fig:exPDP} are marked with circles and listed in Table~\ref{Tab:peaks}. 

\begin{table}[t]
\footnotesize
\renewcommand{\arraystretch}{1.2}
\caption {Peaks Extracted from the PDP Shown in Fig.~\ref{Fig:exPDP}. This PDP is Measured for TX/RX Antenna Positions Shown in Fig.~\ref{setup:fig}.}
\label{Tab:peaks}
\centering
{\begin{tabular}{ccc}
\hline
 Peak \# (i) & $\tau_m(i)$ (ns) &  $P_m(i)$ (dBm)\\\hline
1 & 6.51 & -41.9  \\
2 & 20.83 & -80.4 \\
3 & 30.60 & -72.8\\
4 & 34.50 & -80.1 \\
5 &  45.57 & -78.1 \\
6 &  48.18 & -78.9
\\\hline
\end{tabular}}
\end{table}

Each PDP measured by the sounder contains 2048 samples, and the delay between the samples is $0.651$~ns. The number of peaks extracted from the $m$th measurement is denoted by $N_m$ which is smaller than or equal to the total number of MPCs $N$. Depending on the direction of the TX and the RX antennas, some paths may have small antenna gains as discussed in Section~\ref{sec:Measurements}-A, and these paths may not be detected as they go below the noise floor. For the example PDP in Fig.~\ref{Fig:exPDP}, $N_m=6$ peaks are extracted. The delays of the peaks extracted from the PDP of the $m$th measurement are denoted by $\tau_m(i)$, and the power of the peaks are denoted by $P_m(i)$, where $i=1, \dots, N_m$. 

\begin{rem}
We note that the term $\tau_m(i)$ should not be confused with $\tau_{n,m}$ in~(\ref{eq:delta_tau}). Although both correspond to the $m$th measurement, the former is the \textit{measured} delay of the $i$th peak in the PDP and affected by both the TX/RX antenna direction and the clock drift, whereas the latter depends only on the total distance traveled by the $n$th MPC from the TX antenna to the RX antenna. Furthermore, $\tau_m(i)$ can only take a discrete set of values limited by the delay resolution of the sounder.
\end{rem}

In Fig.~\ref{setup:fig}, the distance between the TX and the RX antennas is set to be 2 m which corresponds to a flight time of 6.67~ns. As the delay resolution of the channel sounder is 0.651~ns, the delay corresponding to the LOS path ($i=1$ in Table~\ref{Tab:peaks}) is measured to be $\tau_m(1)=6.51$~ns which is off by 0.16~ns from the actual delay. This error is reasonable considering the delay resolution of the sounder. A single clock is used in this measurement; therefore, there is no click drift in the measurement.  

In the post-processing, a Hamming window is used to construct the CIRs. As a result, after the power corresponding to a peak is measured, we also take into account two samples adjacent to that peak. According to Friis transmission formula
\begin{equation}
P_{\mathrm{RX}}=P_{\mathrm{TX}}+G_{\mathrm{TX}}+G_{\mathrm{RX}}+ 10 \log_{10} \left( \frac{\lambda}{4 \pi d}\right)^2, 
\end{equation}
we can calculate the theoretical received power corresponding to the LOS path by plugging in the TX power  $P_{\mathrm{TX}}=-10$~dBm and the TX/RX antenna gain $G_{\mathrm{TX}}=G_{\mathrm{RX}}=17$~dBi, and we get $P_{\mathrm{RX}}=-43.4$~dBm. According to Table~\ref{Tab:peaks}, the power of the LOS path is measured to be $P_m(1)=-41.9$~dBm which is larger than the theoretical value only by an amount of 1.5~dB. 



\subsection{Extracting MPCs from Measurements}
The parameters of the MPCs that we extract from the measurements are delay, $\hat{\tau}_n$; path gain, $\hat{\alpha}_n$; AoD,  $\hat{\bm{\theta}}^{\mathrm{AoD}}_n$; and AoA,  $\hat{\bm{\theta}}^{\mathrm{AoA}}_n$. The procedure to extract these parameters is given below.
\begin{itemize}
    \item Step 1: Scan every combination of the azimuth/elevation angles of TX/RX antennas and obtain the PDPs corresponding to those positions. 
    \item Step 2: Extract peaks of each PDP in Step 1 by recording the delay and the received power of the peaks.
    \item Step 3: Correct delays of the extracted peaks considering the clock drift and errors due to antenna rotation (see Section~\ref{sec:Correction}).
    \item Step 4: For each corrected delay, extract only one MPC by selecting the measurement with the highest received power for that delay.  
\end{itemize}
As a result, delays $\hat{\tau}_n$ of the extracted MPCs are the corrected delays. The path gains can be calculated using the following formula

\begin{equation}
\hat{\alpha}_n=-P_{\mathrm{TX}}-G_{\mathrm{TX}}({\bf{0}})-G_{\mathrm{RX}}({\bf{0}})+P_{m_n}(i_n),
\end{equation}
for $n=1,\dots,N$, where $P_{m_n}(i_n)$ is the received power at the $i_n$-th peak of the $m_n$-th measurement. At the $i_n$-th peak, $\tau^{\mathrm{c}}_{m_n}(i_n)=\hat{\tau}_n$, where $\tau^{\mathrm{c}}_{m}(i)$ is the corrected (due to both the clock drift and antenna rotation effects) version of $\tau_{m}(i)$. Furthermore, $\forall m,i$ such that $\tau^{\mathrm{c}}_{m}(i)=\hat{\tau}_n$, we choose the measurement $m_n$ such that $P_{m_n}(i_n) \geq P_{m}(i)$.
We assume that $\bm{\theta}^{\mathrm{AoA}}_n=\bm{\theta}^{\mathrm{RX}}_{m_n}$ and $\bm{\theta}^{\mathrm{AoD}}_n=\bm{\theta}^{\mathrm{TX}}_{m_n}$;
therefore, maximum gains at TX/RX antenna boresight are achieved. Finally, the measurement $m_n$ also specifies the AoD, $\hat{\bm{\theta}}^{\mathrm{AoD}}_n$ and AoA, $\hat{\bm{\theta}}^{\mathrm{AoA}}_n$. Note that the values that $\hat{\tau}_n$ can take are limited by the delay resolution of the sounder. Similarly, the values that $\hat{\bm{\theta}}^{\mathrm{AoD}}_n$ and $\hat{\bm{\theta}}^{\mathrm{AoA}}_n$ can take are limited by the angle resolution  selected. The angle resolution can be improved with the expense of increase in the measurement duration. Step 4 of the MPC extraction procedure is given in detail in Algorithm~\ref{Alg:MpcAlgorithm}.

\algrenewcommand{\algorithmiccomment}[1]{\hskip1.4em$\%$ #1}
\begin{algorithm}[t]
\setstretch{1.2}
\small
\caption{MPC extraction algorithm.}\label{Alg:MpcAlgorithm}
\begin{algorithmic}[1]
\Procedure{ExtractMPCs}{$\tau_m^c(i)$}
\For{$n=1:N$}
\State $\Psi\coloneqq\{(m,i)\mid\ \tau_m^c(i)=\hat{\tau}_n,1\leq m\leq M,1 \leq$ $i \leq N_m\}$
\State $(m_n,i_n)=\underset{(m,i) \in \Psi}{\arg\max}(P_m(i))$
\State $\hat{\tau}_n=\tau_m^c(i)$
\State $\hat{\alpha}_n=-P_{\mathrm{TX}}-G_{\mathrm{TX}}({\bf{0}})-G_{\mathrm{RX}}({\bf{0}})+P_{m_n}(i_n)$
\State $\hat{\bm{\theta}}^{AoD}_n={\bm{\theta}}^{TX}_{m_n}$
\State $\hat{\bm{\theta}}^{AoA}_n={\bm{\theta}}^{RX}_{m_n}$
\State MPC$_n=\{\hat{\bm{\theta}}^{AoD}_n,\hat{\bm{\theta}}^{AoA}_n,\hat{\tau}_n,\hat{\alpha}_n\}$
\EndFor\\
\Return{MPC}
\EndProcedure
\end{algorithmic}
\end{algorithm}

\begin{figure}[t]
\centering
\centerline{\includegraphics[trim=5.2cm 4.0cm 9.5cm 0cm, clip,width=0.49\textwidth]{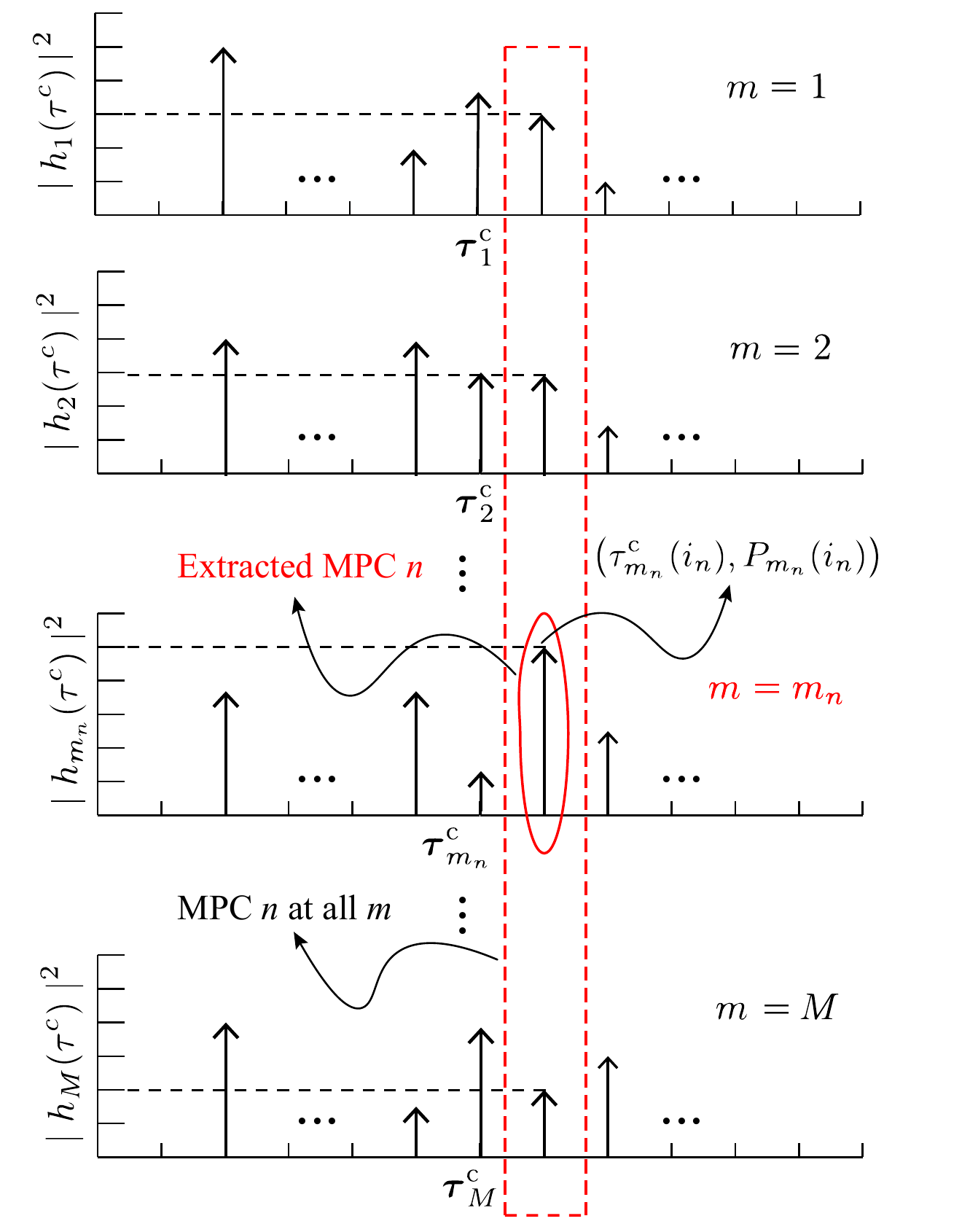}}
\caption{MPC extraction from the PDPs after the clock drift and antenna rotation effects are corrected. The peaks in the dotted box refer to the $n$th MPC in different measurements. Since the strongest peak is obtained at measurement $m_n$, corresponding AoA/AoD are selected to define MPC $n$.}\label{Fig:MPC_extraction}
\end{figure}

MPC extraction process is illustrated in Fig.~\ref{Fig:MPC_extraction} for MPC~$n$. As it can be observed from the figure, peaks in the PDPs, each of which corresponds to the delay of an MPC at different measurements, are now aligned for each delay after the correction steps (Fig.~\ref{Fig:MPC_extraction} shows the peaks at corrected delays whereas Fig.~\ref{Fig:syst} shows the peaks before the corrections). The peaks in different measurements with corrected delays equal to $\hat{\tau}_n$ are shown in the dashed box. Among those, the peak with the greatest power (obtained at measurement $m_n$) is selected as the MPC $n$.

\begin{figure}[t]
\centering
\centerline{\includegraphics[width=0.48\textwidth]{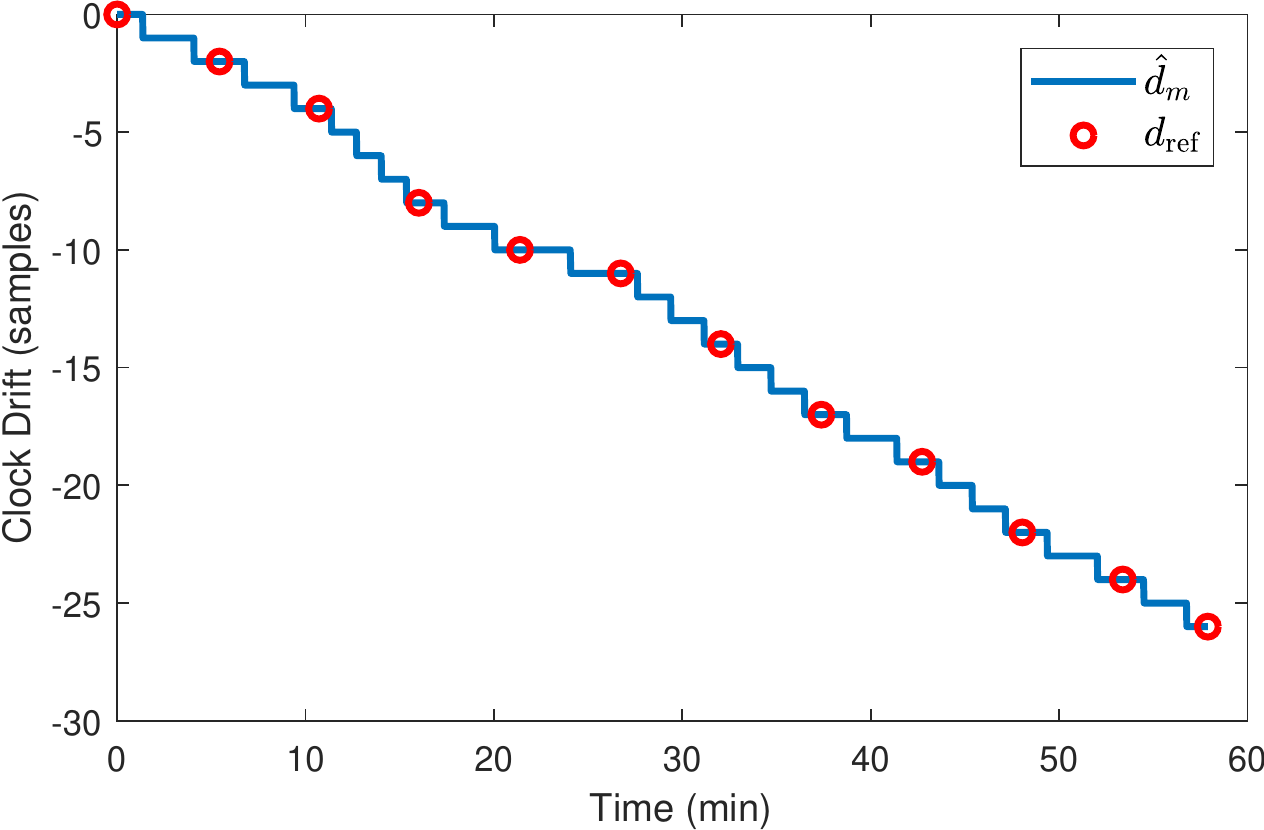}}
\caption{Clock drift from the reference measurements (where the TX and the RX clocks are disconnected) and its fit obtained by linear interpolation followed by quantization. The TX and the RX clocks drift apart from each other by 18~ns in 60~minutes after the disconnection of the synchronization cable.}\label{Fig:drifts}
\end{figure}

\section{Correction of Clock Drift and Delay Error due to Antenna Rotation}
\label{sec:Correction}
As discussed in Section~\ref{sec:Measurements}-A, there are two major problems for accurate extraction of the MPCs: 1) the drift between the TX and the RX clocks, and 2) the error in the delay term due to the rotation of the TX/RX antennas. These problems may result in the phenomenon which we will refer to as the \textit{missed} and \textit{ghost} MPCs. A missed MPC is an MPC which exists in the measurements performed with a single clock; however, it is missed in the separate-clock measurements due to the miscalculation of some of the MPC parameters such as delay. On the other hand, a ghost MPC is an MPC which does not exist in reality; however, it is extracted from the separate-clock measurements by mistake. In this section, we will illustrate the effect of the errors mentioned above in extracting the MPCs and propose an algorithm to fix these errors.\looseness=-1

\subsection{Tracking and Correction of the Clock Drift}

To be able to correct the clock drift, the drift is required to be tracked. To achieve this, we use the delay corresponding to a strong LOS path that is recorded at each reference measurement. Let $\tau_{\mathrm{ref,}k}$ be the delay of the LOS path at the $k$th reference measurement. If the measurements were performed in a stationary environment and when there was no drift between the TX and the RX clocks (i.e., when the synchronization cable was connected), then $\tau_{\mathrm{ref,}k}$ would be the same for all $k=1,\dots,K$. However, for the separate-clock measurements, since the TX and the RX clocks drift apart from each other as the time passes, $\tau_{\mathrm{ref,}k}$ value changes with $k$. 

For the first reference measurement, timing alignment is performed so that $\tau_{\mathrm{ref},1}$ corresponds to the delay of the LOS path and hence, $d_{\mathrm{ref},1}=0$. First, the relative delay (or the drift) of the LOS path at each reference measurement is calculated as $d_{\mathrm{ref},k}=\tau_{\mathrm{ref},k}-\tau_{\mathrm{ref},1}$, for $2<k<K$. An example clock drift for a 1-hour long measurement is shown in Fig.~\ref{Fig:drifts}, where $d_{\mathrm{ref}}$'s are indicated by the circular red markers and the drift is shown in samples.  Next, the drift at measurement $m$ is estimated by interpolating the two consecutive $d_\mathrm{ref}$ values around measurement $m$ linearly and quantizing to the nearest integer.  
Once the drift estimate $\hat{d}_m$ is found for all $m$,  correction of clock drift (CCD) is performed as follows:
\begin{equation}
\tau^{c'}_m(i)=\tau_m(i)-\hat{d}_m
\end{equation}
for all $i$ and $m$. 

Once the clock drift is corrected, the next step is the correction of delay error due to antenna rotation (CDEDAR) to obtain $\tau^{c}_m(i)$ which is explained in the next subsection. The delay correction in overall is thus as follows:
\begin{equation}
\tau_m(i) \xrightarrow{CCD} \tau^{c'}_m(i) \xrightarrow{CDEDAR} \tau^c_m(i).
\end{equation}


\begin{figure}[!t]
	\centering
	\begin{subfigure}{\columnwidth}
	\centering
	\includegraphics[width=0.49\textwidth]{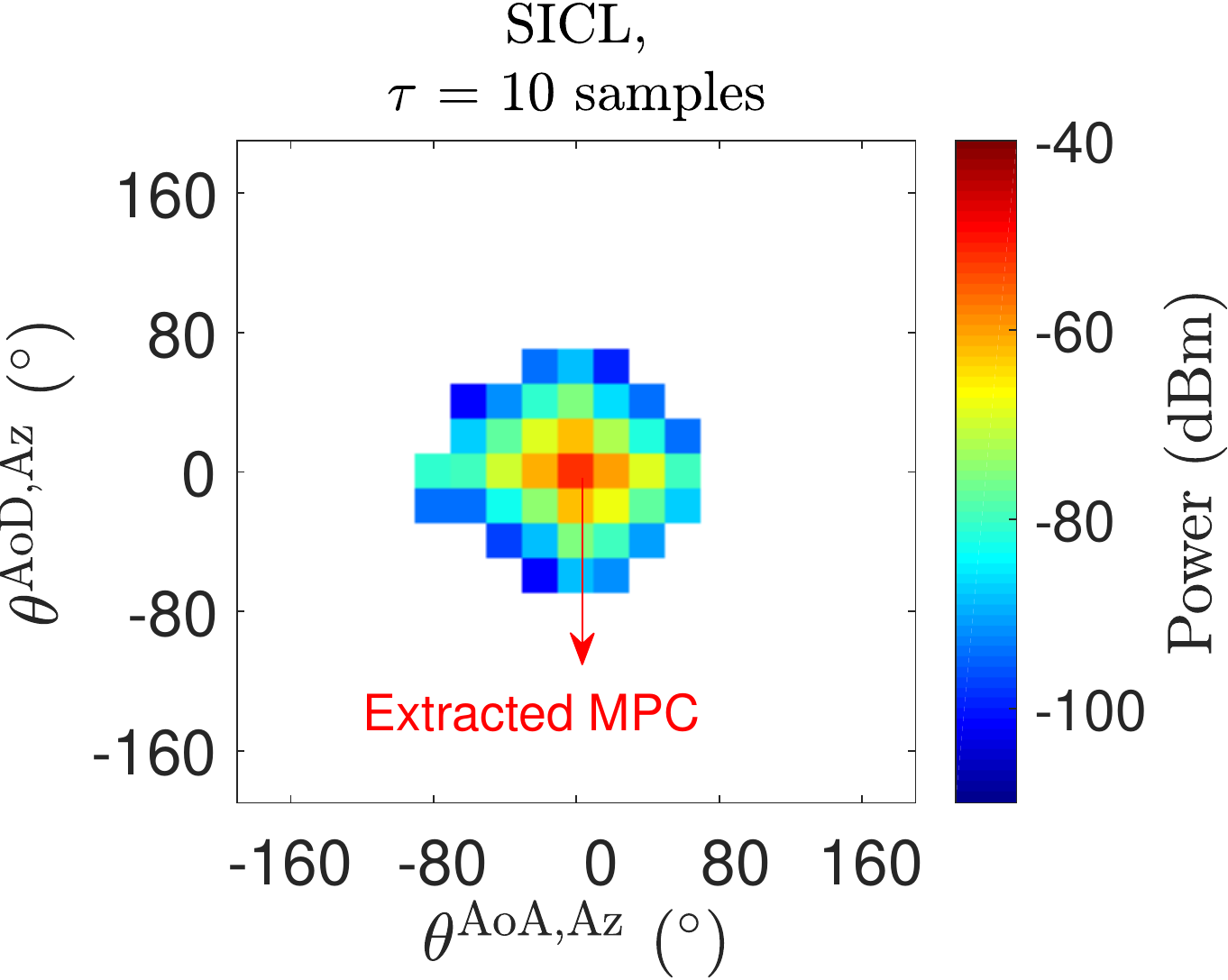}
	\hfill
	\includegraphics[width=0.49\textwidth]{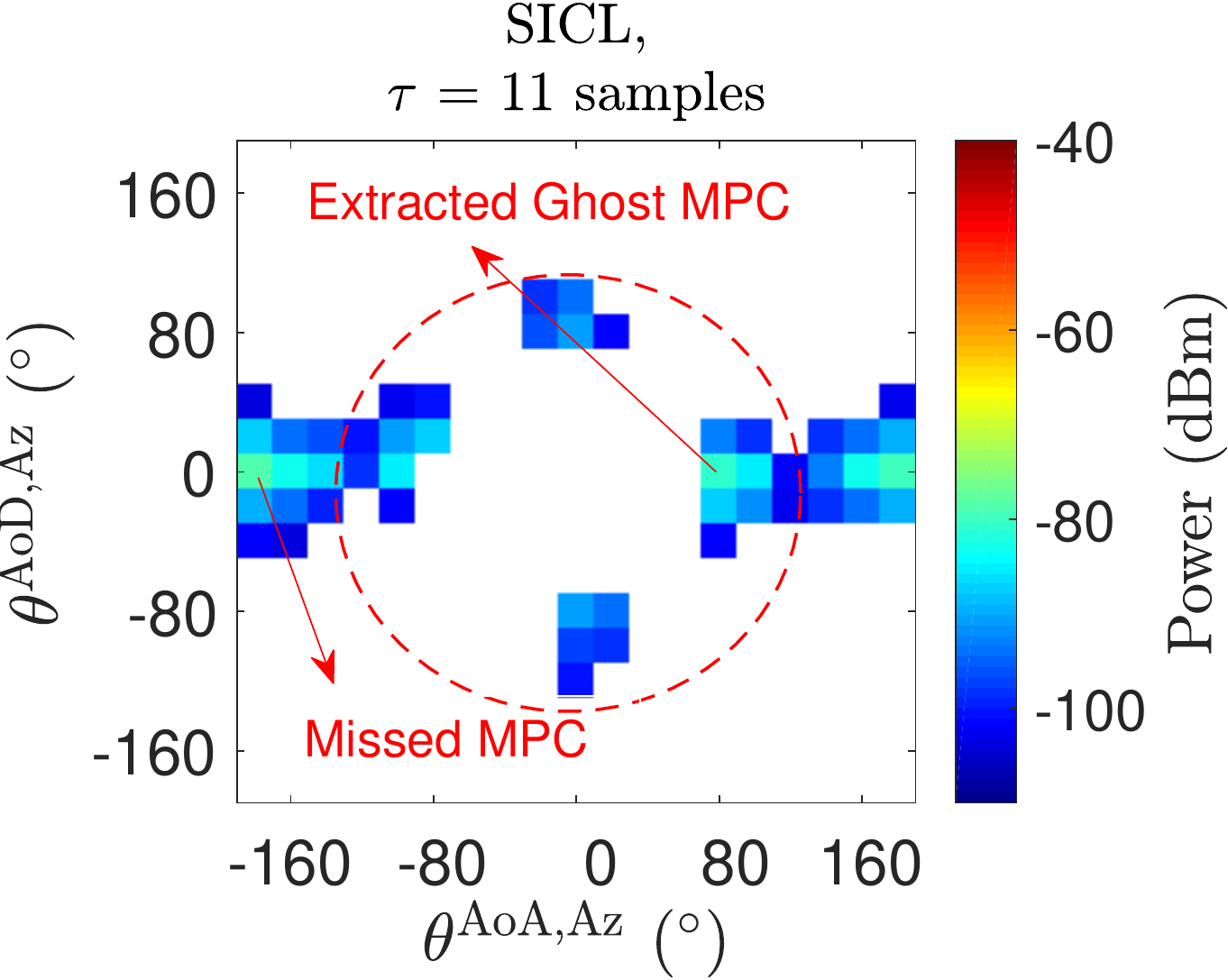}
	\caption{}
    \end{subfigure}	

	\begin{subfigure}{\columnwidth}
	\centering
	\includegraphics[width=0.49\textwidth]{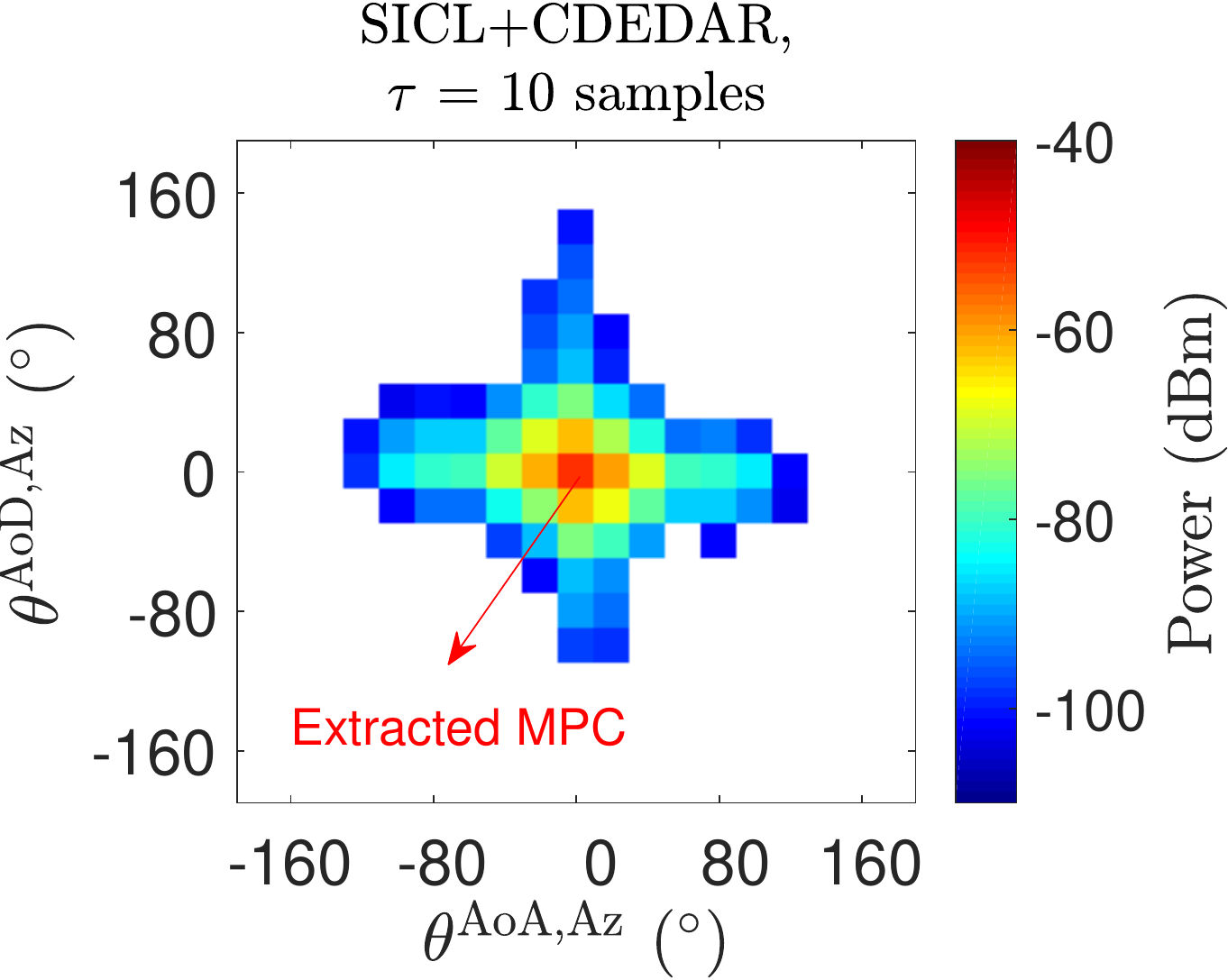}
	\hfill
		\includegraphics[width=0.49\textwidth]{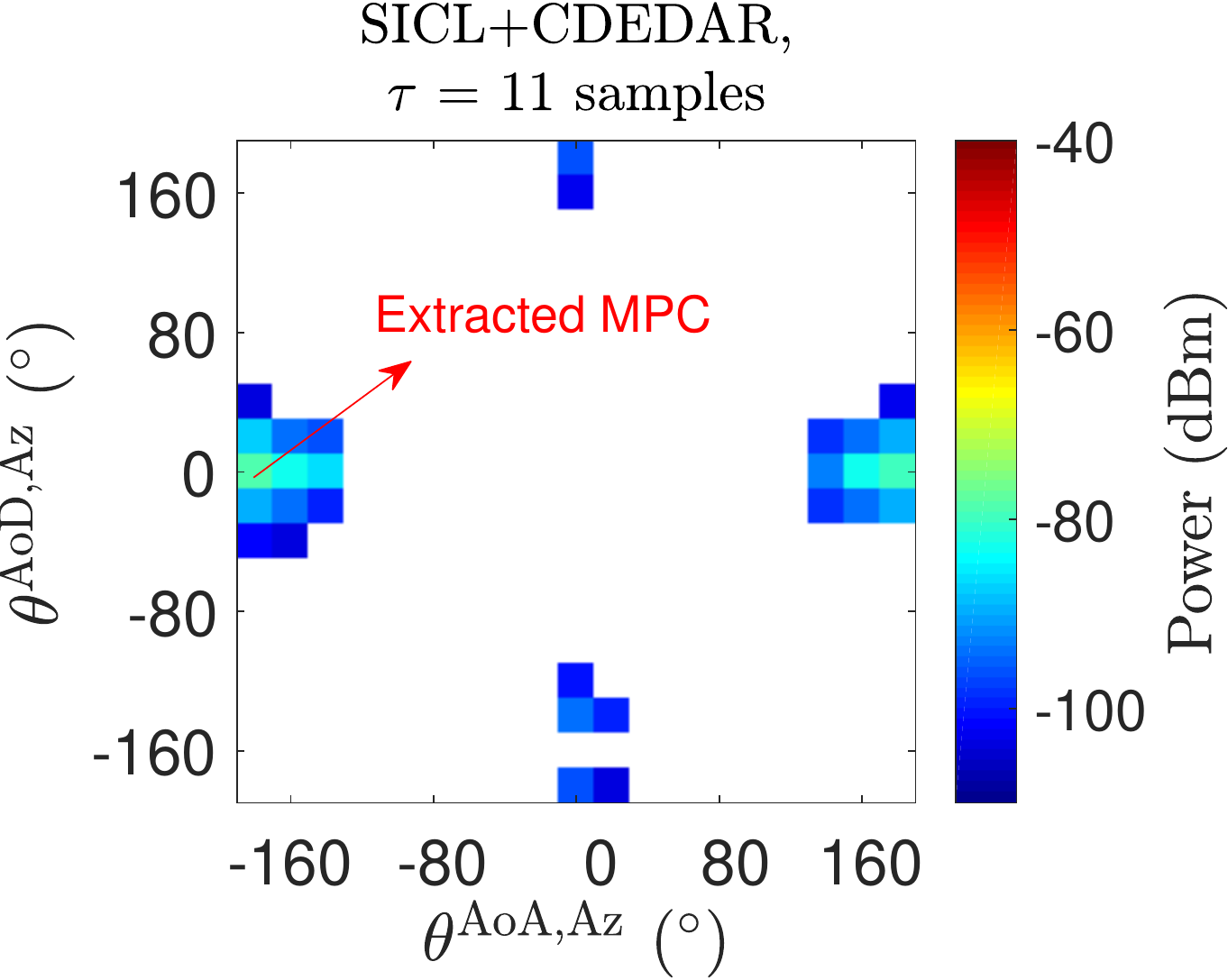}
	\caption{}
    \end{subfigure}
     \caption{Received power in the AoA azimuth-AoD azimuth plane for all the detected peaks with delays of 10 samples and 11 samples for the (a) raw single-clock measurement, and (b) antenna error corrected single-clock measurement. TX $El$ = $0^{\circ}$, RX $El$ = $20^{\circ}$. }
     \label{Fig:MPCs-single}
\end{figure}

\subsection{Correction of the Delay Error Due to Antenna Rotation}

The rotation of the TX/RX antennas causes the error, $\Delta\tau_{n,m}$, in the delay of the measured peaks. Due to the rotation of antennas, the distance traveled by the same MPC can differ by up to 40~cm, as shown in Fig.~\ref{Fig:syst}. Thus, the delay $\tau_{n,m}$ measured for the $n$th MPC at measurement $m$ can deviate from $\tau_{n,\mathrm{min}}$ by up to $\approx1.33$~ns which is more than twice of our delay resolution. Since the peak that has the greatest received power, $P_{m_n}(i_n)$, among the other peaks at delay $\hat{\tau}_n$ qualifies as an MPC (see Section~\ref{sec:Measurements}-C), the error in the delay of the peaks may cause missed/ghost MPCs. 


\algrenewcommand{\algorithmiccomment}[1]{\hskip1.4em$\%$ #1}
\begin{algorithm}[t]
\setstretch{1.2}
\small
\caption{Algorithm for CCD and CDEDAR}\label{Alg:MpcCorrection}
\begin{algorithmic}[1]
\Procedure{CorrectErrors}{$\tau_m(i), P_m(i), \tau_{\mathrm{ref,}k}$}
\\ 
\Comment{CCD}
\For{$k=1:K$}
\State $d_{\mathrm{ref},k}=\tau_{\mathrm{ref},k}-\tau_{\mathrm{ref},1}$
\EndFor
\State fit a linear model to each two successive $d_{\mathrm{ref,}k}$ values

\quad to estimate $d_m$
\algrenewcommand{\algorithmiccomment}[1]{\hskip2.1em$\triangleright$ #1}
\For{$m = 1:M$}
\For{$i = 1:N_m$}
\State $\tau^{c'}_m(i)= \tau_m(i)-\hat{d}_m$
\EndFor
\EndFor
\algrenewcommand{\algorithmiccomment}[1]{\hskip1.4em$\%$ #1}
\\ \Comment{CDEDAR}
\Repeat
\For{$m = 1:M$}
\For{$i = 1:N_m$}
\State find the set of measurements in the 

\quad \quad \quad \quad \quad \quad neighborhood of $m$: $M^\prime$
\State $\Omega\coloneqq\{(m^\prime,i^\prime)\mid m^\prime \in M^\prime, \tau_{m^\prime}(i^\prime)=\tau^{c'}_m(i)\lor$

\quad \quad \quad \quad \quad \quad $\tau_{m^\prime}(i^\prime)=\tau^{c'}_m(i)-1\}$
      \If{peak with max power in $\Omega$ has a delay of
      
      \quad \quad \quad \quad \quad \quad $\tau^{c'}_m(i)-1$}
        \State $\tau^{c'}_m(i)  \gets \tau^{c'}_m(i)-1$ 
      \EndIf
\State $\tau^{\mathrm{c}}_{m}(i) = \tau^{c'}_{m}(i)$
\EndFor
\EndFor
\Until{converge (see Remark 2)}\\
\Return{$\tau^{\mathrm{c}}_{m}(i)$}
\EndProcedure
\end{algorithmic}
\end{algorithm}

The problem with the rotation of the antennas is illustrated in Fig.~\ref{Fig:MPCs-single}(a). The figure at left shows the power of the peaks received at 10 samples delay from the single-clock (SICL) measurements when the elevation angles are fixed at $0^{\circ}$ and $20^{\circ}$ for the TX and the RX antennas, respectively. We expect to see more peaks with significant amount of power at the surrounding TX/RX azimuth angles ($Az$); however, they are missing. Since there is no clock drift in SICL measurements, the errors here are only due to antenna rotation. The distance traveled by the components that actually follow the same path changes with the rotation of the antennas. Therefore, some of the peaks, which are supposed to be received at 10 samples delay, appear at a later delay of 11 samples (see the peaks in the dashed circle in the figure at right).
As a result, a ghost MPC is detected at TX $Az$ = $0^{\circ}$ and RX $Az$ = $80^{\circ}$ for the 11 samples delay. In addition, the actual MPC for the 11 samples delay at TX $Az$ = $0^{\circ}$ and RX $Az$ = $-168^{\circ}$ is shadowed by the ghost MPC and missed. 

Based on the discussion above, the delay of the received peaks should also be corrected against the errors due to the rotation of the antennas. 
This is done as follows. After CCD is applied, $\tau^{c'}_{m}(i)$ is the delay (in samples) of the $i$th peak at measurement $m$. First, all the peaks detected at a delay of $\tau^{c'}_{m}(i)$ or $\tau^{c'}_{m}(i)-1$ samples in the neighborhood of the $m$th measurement are listed.   
A measurement is considered within the neighborhood of the other measurement when the difference between the measurement angles in any plane is no more than 20 degrees. Then, if the delay of the peak in the list with maximum power is equal to $\tau^{c'}_{m}(i)-1$, the corrected delay $\tau^c_m(i)$ is updated as $\tau^{c'}_{m}(i)-1$; otherwise, $\tau^c_m(i)$ becomes $\tau^{c'}_{m}(i)$. This process is done for all $i$ and $m$, and repeated several times until there is no change in the delay of any peak. The overall procedure for the proposed CCD and CDEDAR is given in Algorithm~\ref{Alg:MpcCorrection}. \looseness=-1

\begin{rem}
In Algorithm 2, during the implementation of CDEDAR, the delay of a peak can be changed at most two times through the iterations because the error in the delay due to the antenna rotation is limited to 2 samples. Algorithm~\ref{Alg:MpcCorrection} converges when there is no change in any $\tau_m(i)$ in the last two successive iterations.
\end{rem}

\begin{table}
\vspace{-1mm}
\footnotesize
\renewcommand{\arraystretch}{1.2}
	\caption{Parameters of the MPCs Extracted from the Single-Clock and Separate-Clock Measurements. TX $El$ = $0^{\circ}$, RX $El$ =~$20^{\circ}$.}
	\label{tab:MPCs10-11}
	\begin{subtable}[t]{8cm}
		\centering
		\begin{tabular}{m{2.5cm}cccc}
\hline
\\[-1.2em]
 Source& $\theta^{\mathrm{AoD,Az}}$ ($^{\circ}$) & $\theta^{\mathrm{AoA,Az}}$ ($^{\circ}$) & Power (dBm)\\\hline
SICL& 0 & 0 &   -52.34 \\
SICL+CDEDAR& 0 & 0 &  -52.34\\
SECL& -168 & 0 &  -76.10\\
SECL+CCD& 0 & 0 &  -52.33\\
SECL+CCD\&CDEDAR&  0 & 0 &  -52.33
\\\hline
\end{tabular}
		\caption{10 samples}\label{tab:MPCs10-11-a}
	\end{subtable}
	\quad
	\begin{subtable}[t]{8cm}
		\centering
		\begin{tabular}{m{2.5cm}cccc}
\hline
\\[-1.2em]
Source & $\theta^{\mathrm{AoD,Az}}$ ($^{\circ}$) & $\theta^{\mathrm{AoA,Az}}$ ($^{\circ}$) & Power (dBm)\\\hline
SICL & 0 & 80 &  -80.53 \\
SICL+CDEDAR& 0 & -168 & -78.93 \\
SECL& -40 & 80 &  -79.86\\
SECL+CCD  & 20 & 20 & -72.24 \\
SECL+CCD\&CDEDAR& 0 & -168 & -78.77
\\\hline
\end{tabular}
		\caption{11 samples}\label{tab:MPCs10-11-b}
	\end{subtable}
	\label{table:1}
\end{table}

The received power in the AoA azimuth$-$AoD azimuth plane after applying the above procedure is shown in Fig.~\ref{Fig:MPCs-single}(b). The delay of the peaks which appear at 11 samples delay is corrected, and some of the peaks are moved next to the other peaks with 10 samples delay. Consequently, the ghost MPC is replaced by the actual MPC. 
After CDEDAR is applied, the parameters of the MPC extracted at 11 samples delay are $\theta^{\mathrm{AoD,Az}} = 0^\circ$ and $\theta^{\mathrm{AoA,Az}} = -168^\circ$. Parameters of the MPCs extracted from the raw SICL measurement, SICL measurement after CDEDAR, raw separate-clock (SECL) measurement, SECL measurement after CCD, and SECL measurement after CCD and CDEDAR are presented in Table~\ref{tab:MPCs10-11}(a) and Table~\ref{tab:MPCs10-11}(b), respectively for delays of 10 samples and 11 samples.



\begin{figure}[!t]
	\centering
	\begin{subfigure}{\columnwidth}
	\centering
	\includegraphics[width=0.49\textwidth]{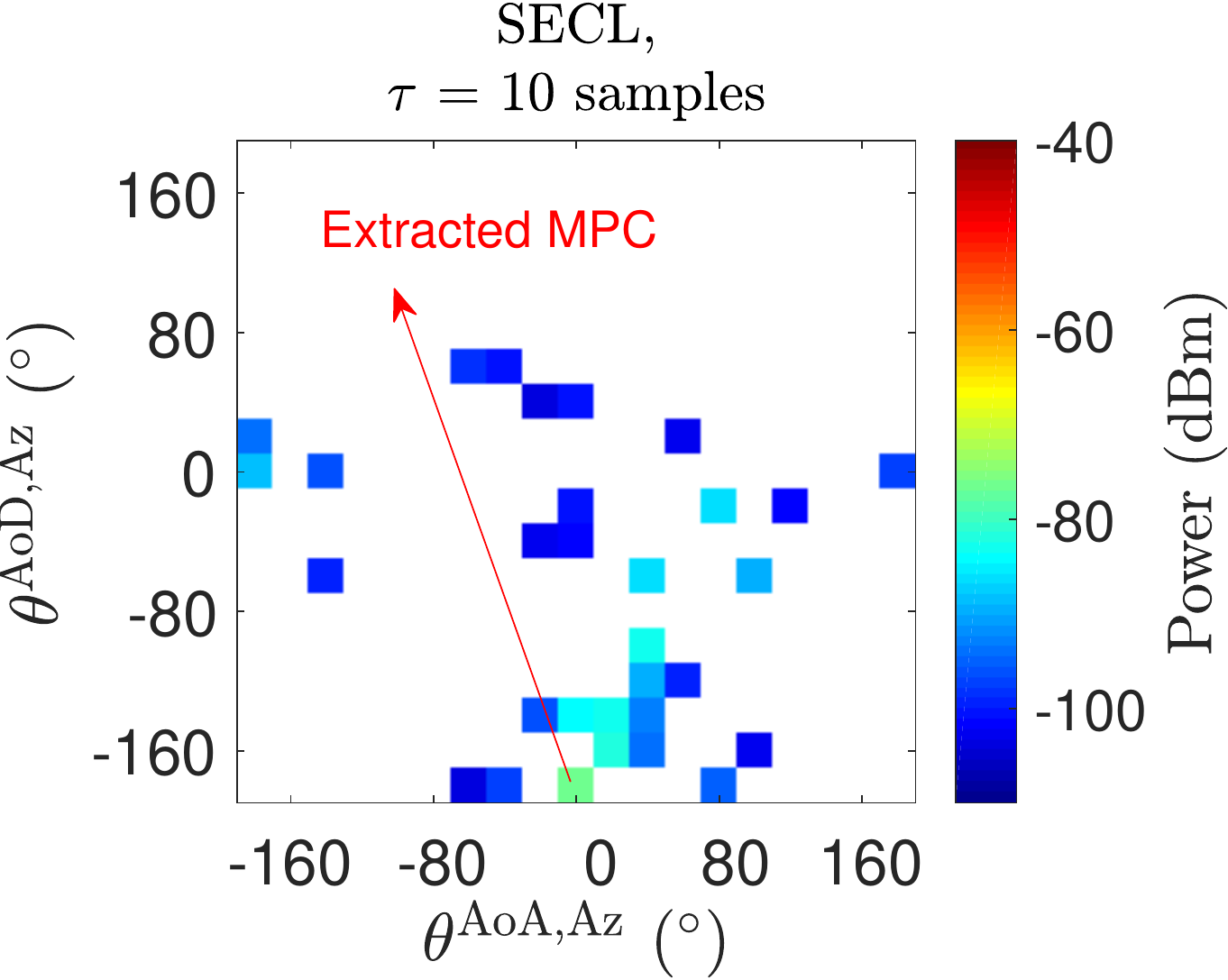}
	\hfill
	\includegraphics[width=0.49\textwidth]{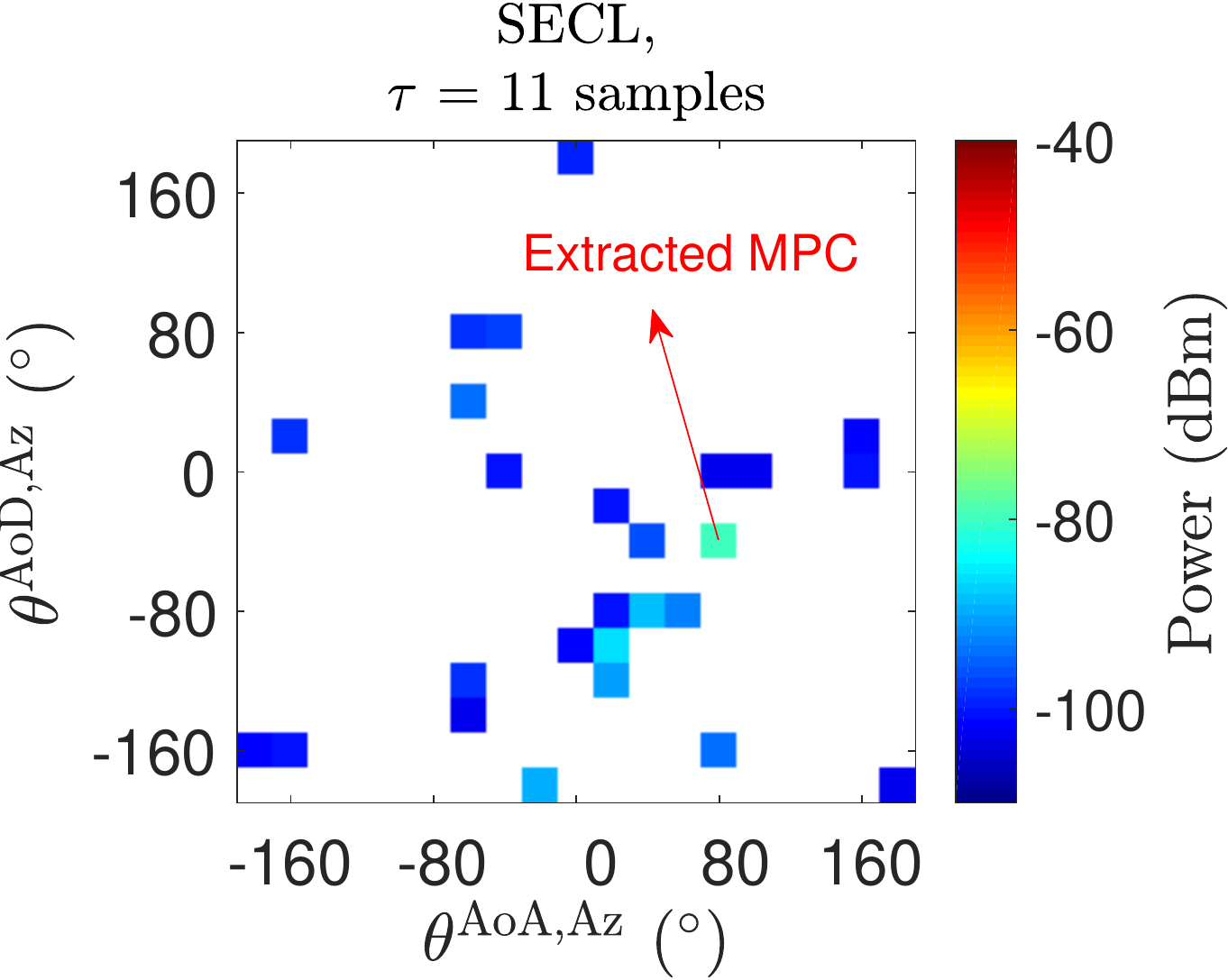}
	\vspace{-5mm}
	\caption{}
    \end{subfigure}
    
    \begin{subfigure}{\columnwidth}
	\centering
	\includegraphics[width=0.49\textwidth]{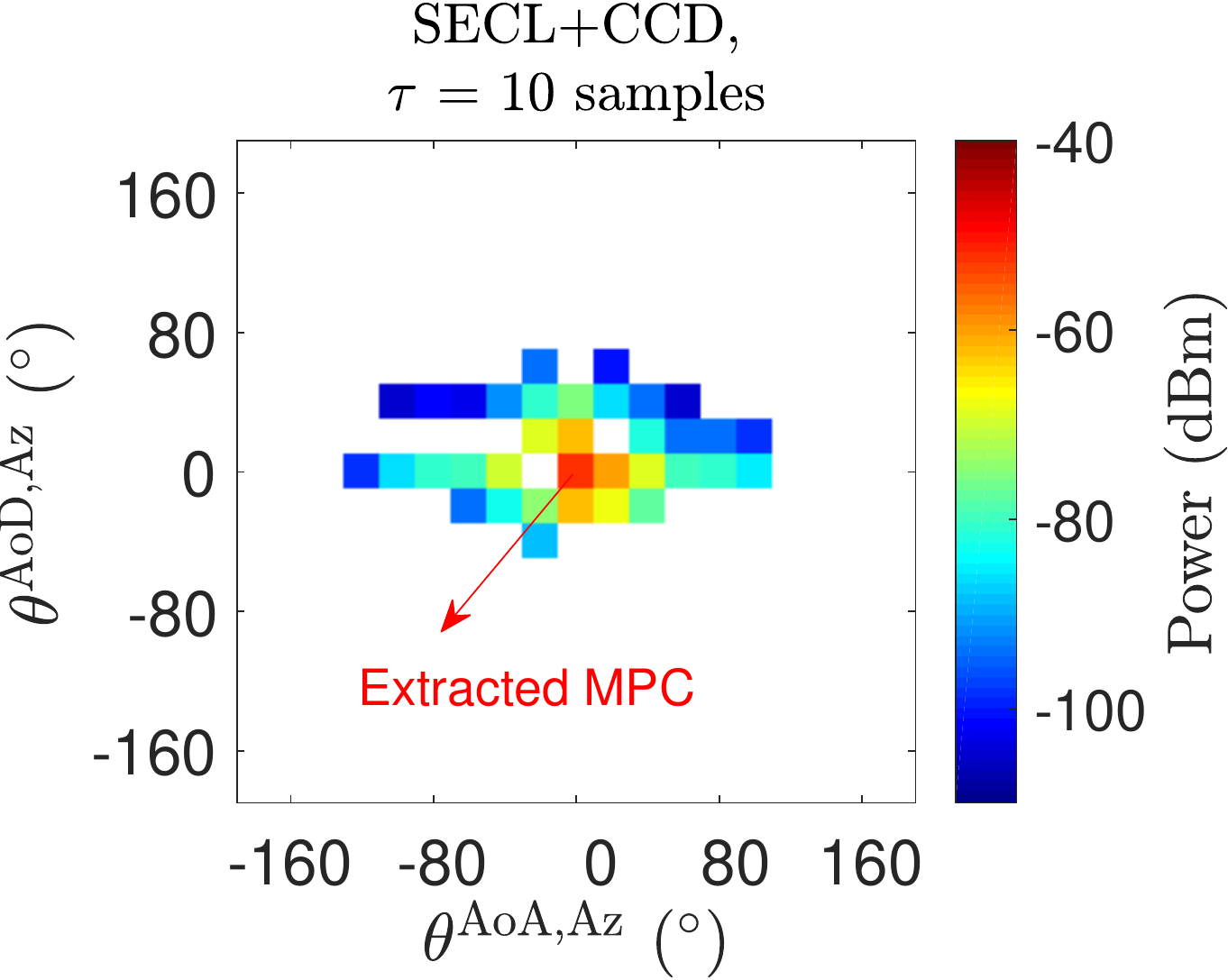}
	\hfill
	\includegraphics[width=0.49\textwidth]{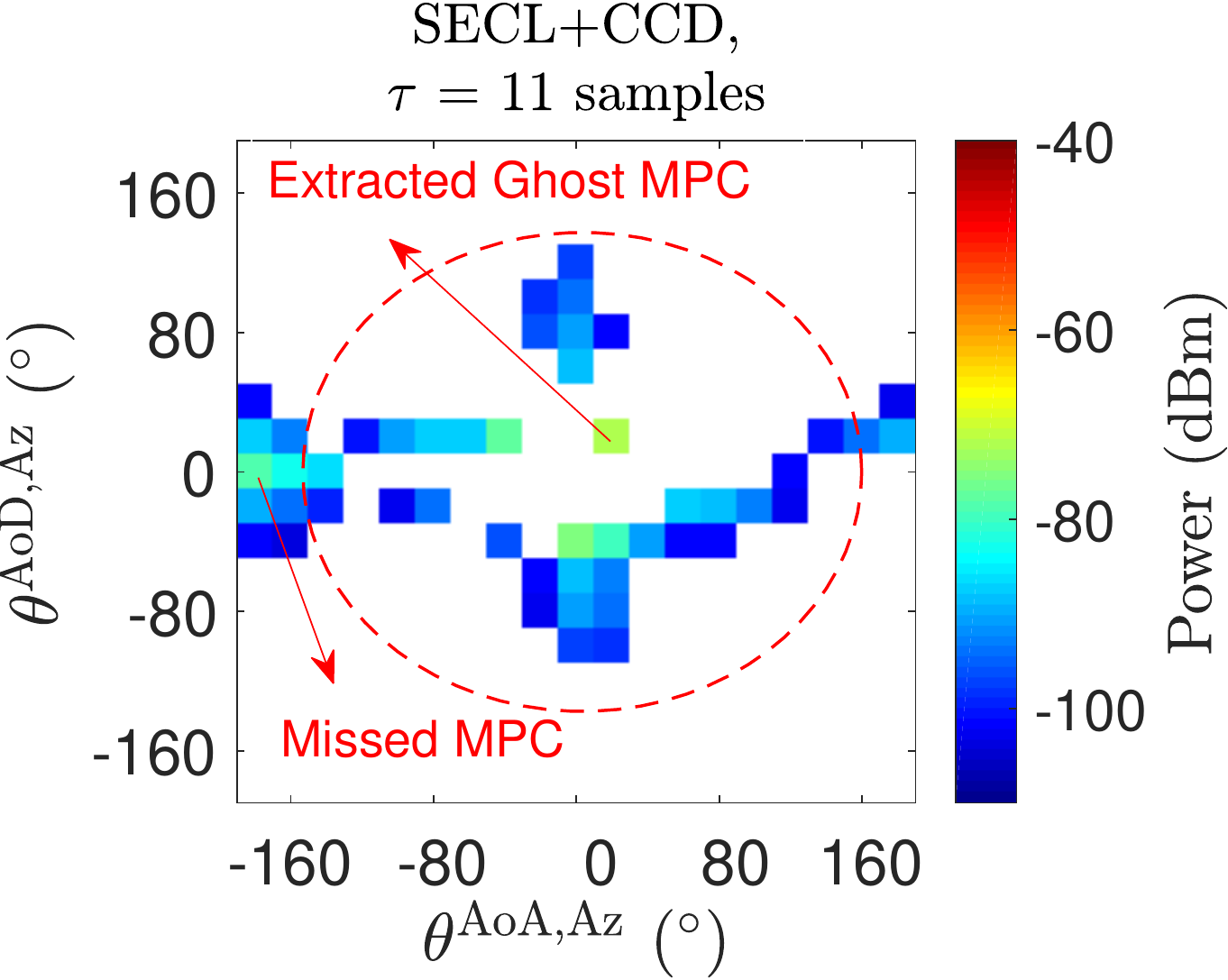}
	\vspace{-5mm}
	\caption{}
    \end{subfigure}
    
	\begin{subfigure}{\columnwidth}
	\centering
	\includegraphics[width=0.49\textwidth]{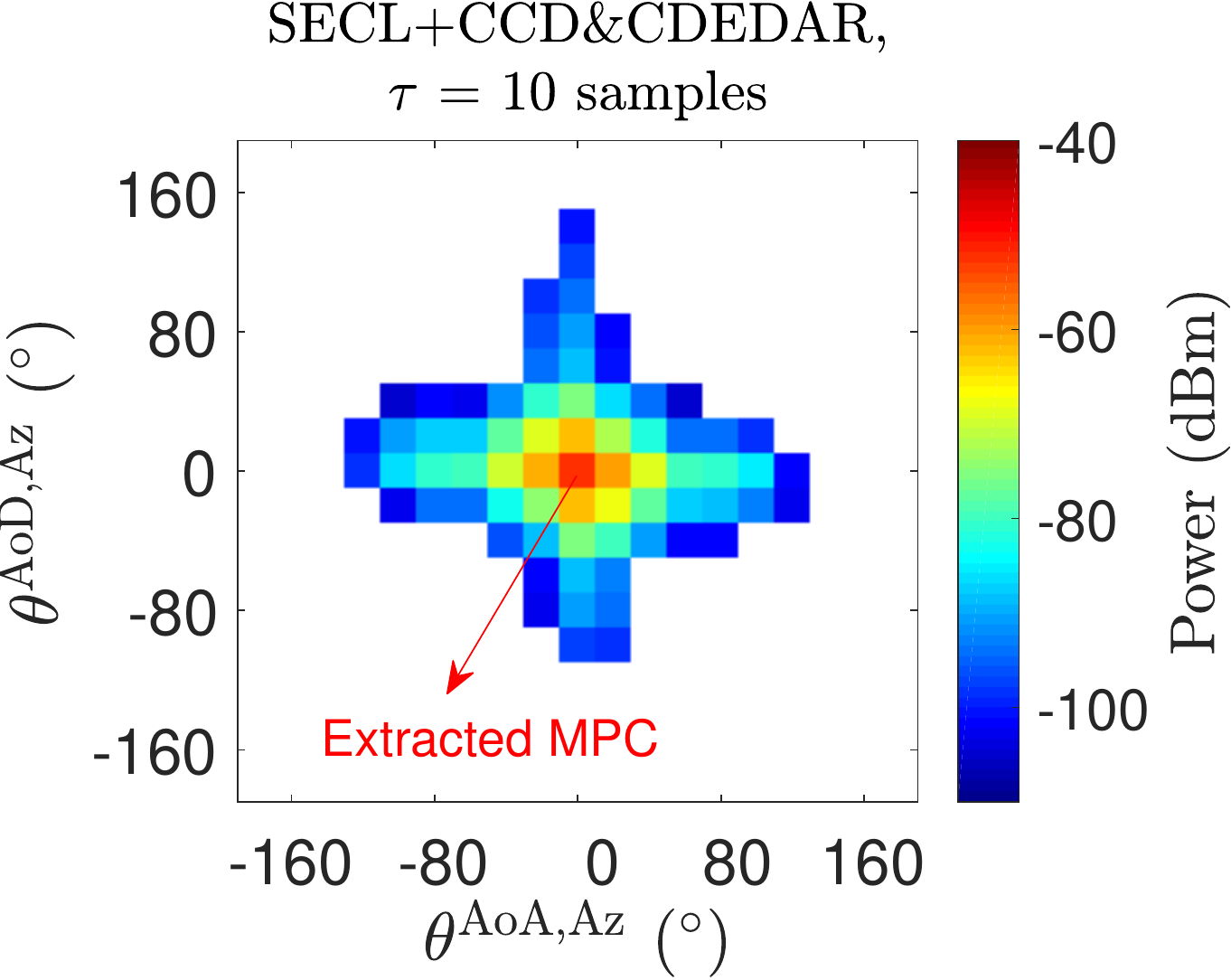}
	\hfill
	\includegraphics[width=0.49\textwidth]{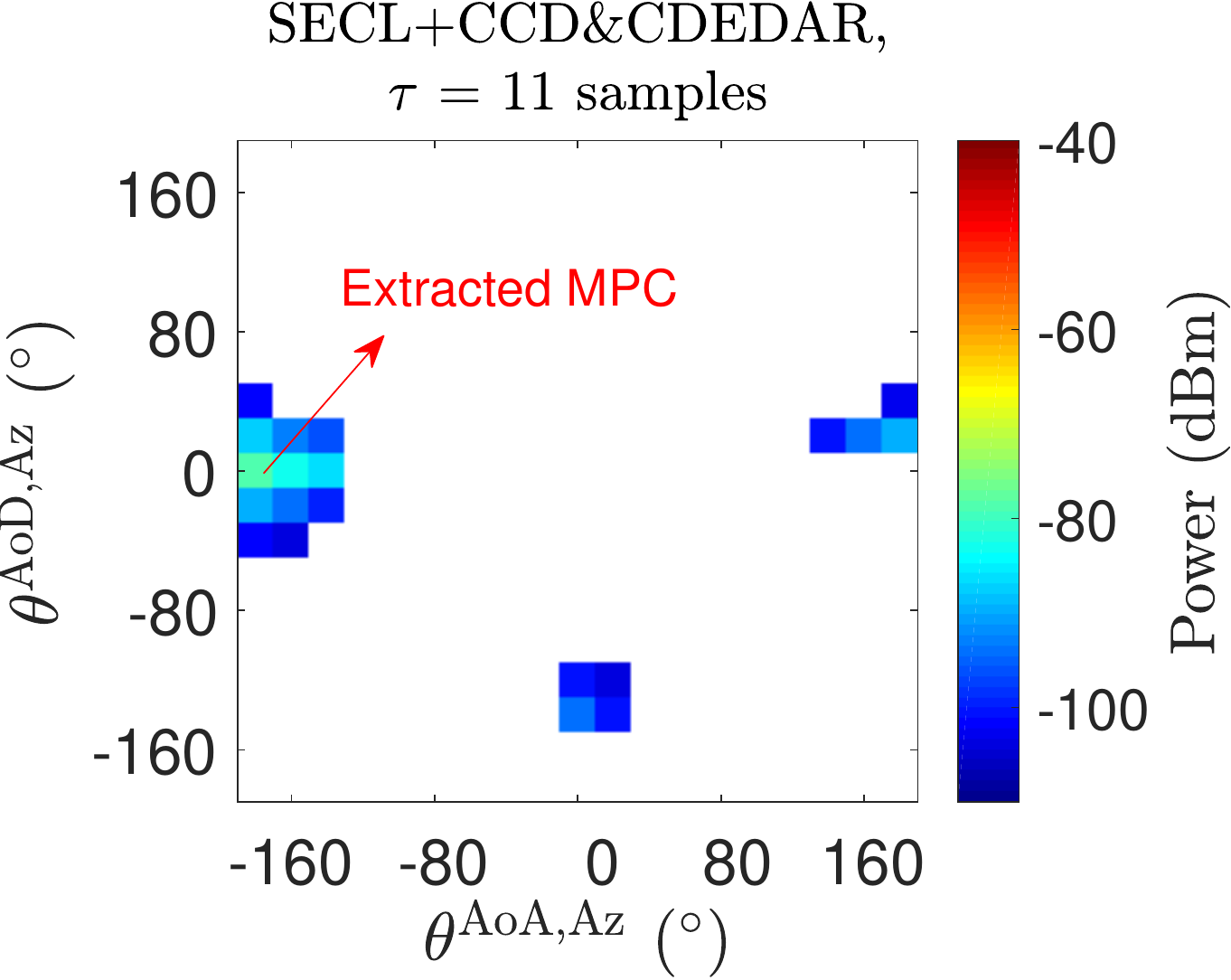}
	\vspace{-5mm}
	\caption{}
    \end{subfigure}	
     \caption{Received power in the AoA azimuth-AoD azimuth plane for all the detected peaks with delays of 10 samples and 11 samples for the (a) raw separate-clock measurement, (b) clock drift corrected separate-clock measurement, and (c) both the clock drift and antenna error corrected separate-clock measurement. TX $El$ = $0^{\circ}$, RX $El$ = $20^{\circ}$. }
     \label{Fig:MPCs-drifting}
\end{figure}

\subsection{Correcting Combined Effects of the Clock Drift and the Antenna Rotations}
Power in the AoA azimuth$-$AoD azimuth plane for the SECL measurements is plotted in Fig.~\ref{Fig:MPCs-drifting}(a)-(c). It is clear from the comparison of Fig.~\ref{Fig:MPCs-drifting}(a) and Fig.~\ref{Fig:MPCs-single}(b), it may be completely incorrect to extract MPCs from the raw SECL measurements. CCD provides a significant improvement towards the accurate detection of the MPCs, as shown in Fig.~\ref{Fig:MPCs-drifting}(b); however, the missed/ghost MPC problem still persists. Fig.~\ref{Fig:MPCs-drifting}(c) shows the results for the SECL measurement after both the errors corrected, i.e., after applying CCD and CDEDAR. As it can also be verified from the second and fifth rows in Table~\ref{tab:MPCs10-11}(a) and Table~\ref{tab:MPCs10-11}(b), after the correction steps, the MPC parameters obtained for the SECL measurements are very close to those of the corrected SICL measurements.
\begin{figure*}[t]
	\begin{subfigure}{0.49\textwidth}
	\centering
	\includegraphics[width=\textwidth]{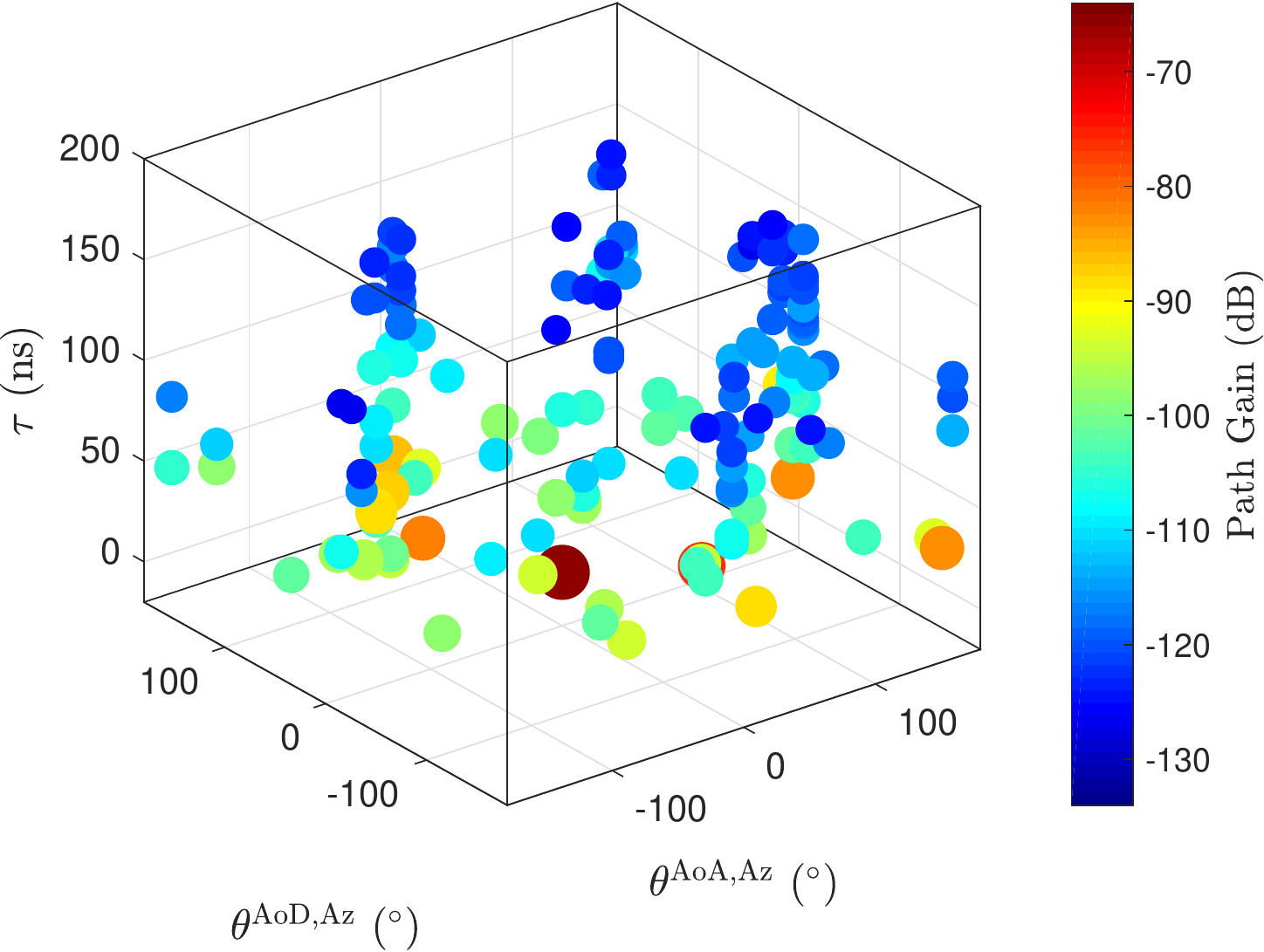}
	\caption{SICL After CDEDAR}
    \end{subfigure}			
	\begin{subfigure}{0.49\textwidth}
	\centering
    \includegraphics[width=\textwidth]{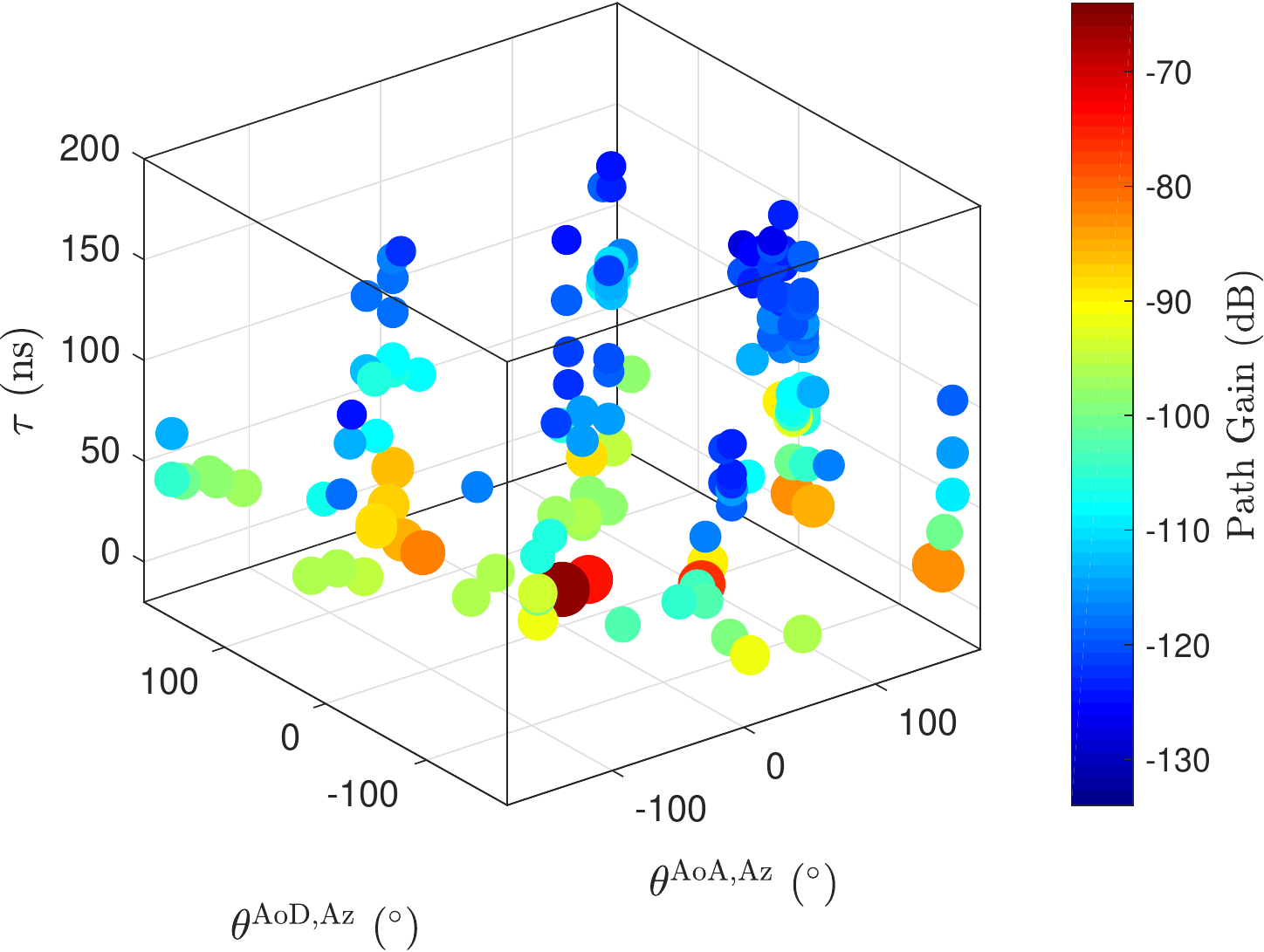}
	 \caption{SECL Raw}
     \end{subfigure}
     \begin{subfigure}{0.49\textwidth}
	\centering
	\includegraphics[width=\textwidth]{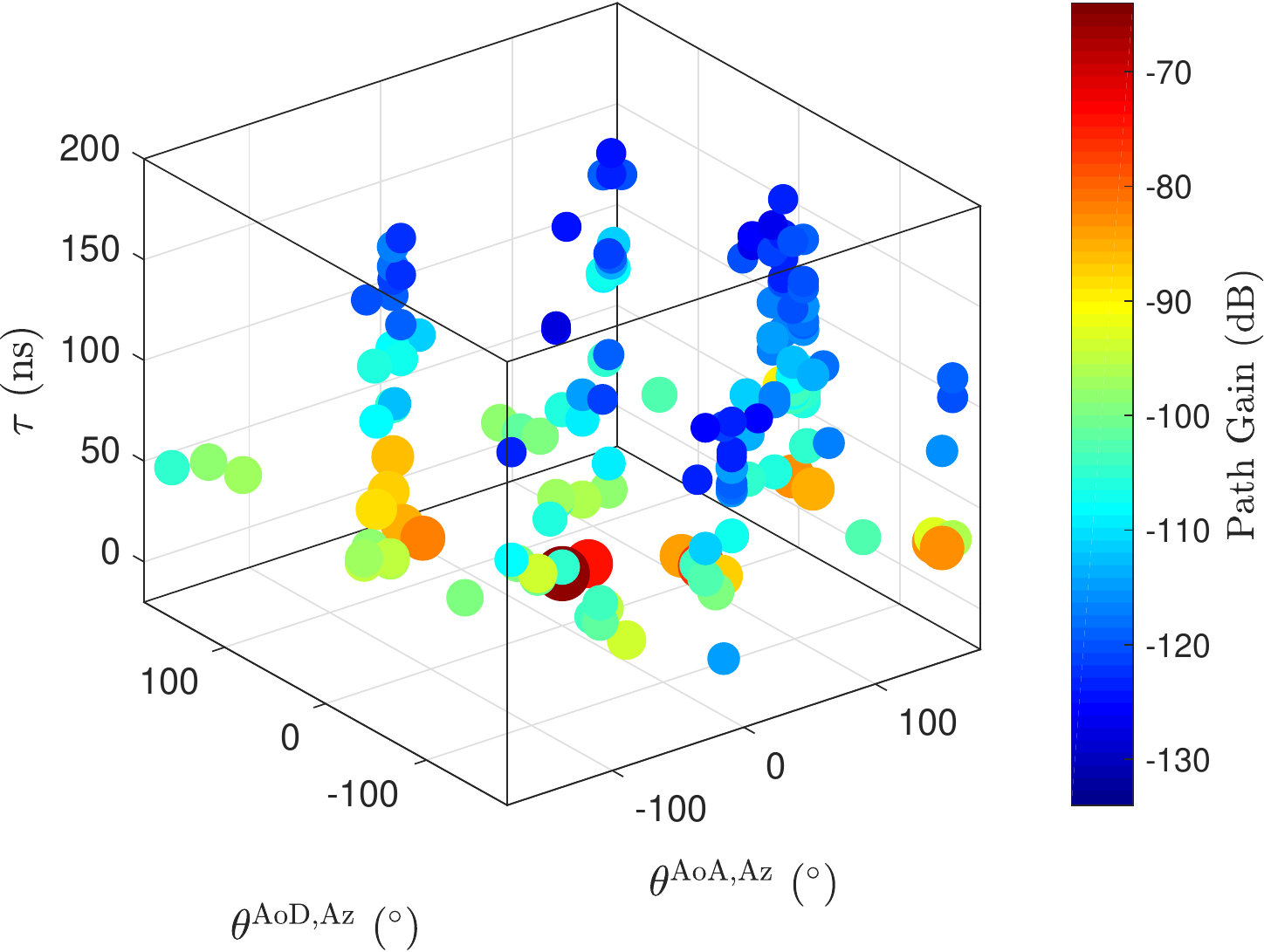}
	\caption{SECL After CCD}
    \end{subfigure}
    \hspace{2mm}
	\begin{subfigure}{0.49\textwidth}
	\centering
    \includegraphics[width=\textwidth]{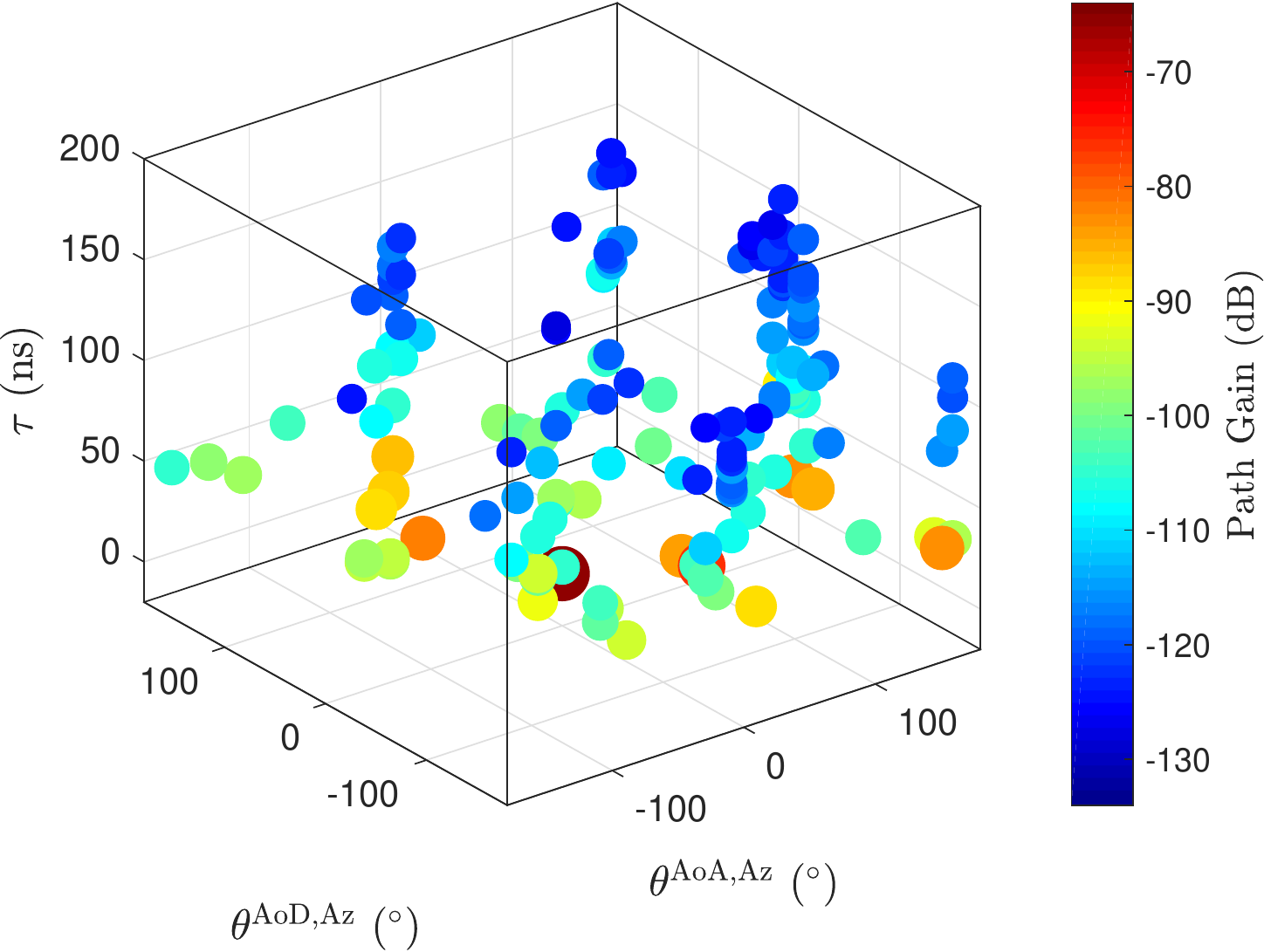}
	 \caption{SECL After CCD \& CDEDAR}
     \end{subfigure}
     \caption{Power in the delay-AoA azimuth-AoD azimuth plane (when TX $El = 0^\circ$, RX $El = 0^\circ$) for the (a) single-clock measurement after antenna rotation errors corrected, (b) separate-clock raw measurement, (c) separate-clock measurement after clock drift corrected, and (d) separate-clock measurement after both the clock drift and delay error due to antenna rotations corrected. Greater power values are represented by larger circles.}
     \label{Fig:DelayPower}
\end{figure*}

To have a better understanding of the proposed MPC extraction algorithm, the results can be visualized for the SICL measurement after CDEDAR, and SECL raw and post-processed measurements in the delay-AoA azimuth-AoD azimuth plane. Fig.~\ref{Fig:DelayPower} shows the power of all the extracted MPCs from the specified measurements for all delays and TX/RX $Az$ where both the TX $El$ and RX $El$ are fixed at
$0^{\circ}$. Since we are more interested in the MPCs with higher powers, those MPCs are represented by larger circles. Fig.~\ref{Fig:DelayPower}(a) shows the MPCs extracted from the SICL measurement after CDEDAR is applied. It can be visually verified that the MPCs from the SECL measurement after implementing CCD and CDEDAR, shown in Fig.~\ref{Fig:DelayPower}(d), demonstrate a noticeably better match to those shown in Fig.~\ref{Fig:DelayPower}(a) when compared with the MPCs from the SECL measurement where only the CCD is applied (see Fig.~\ref{Fig:DelayPower}(c)). The improvement is much more substantial when the MPCs shown in  Fig.~\ref{Fig:DelayPower}(a) are compared with those of the SECL raw measurements (see Fig.~\ref{Fig:DelayPower}(b)). A quantitative analysis of the accuracy of the MPCs that are extracted from the SECL measurements after the correction steps will be made in Section~\ref{Sec:Results}. Before proceeding with the experimental results, we introduce the Hungarian algorithm based MPC matching method.

\subsection{MPC Matching Problem}
As the SICL measurements are not subject to clock drift, we treat the extracted MPCs after applying CDEDAR to SICL measurements as our ground truth. The contribution of performing CCD and CDEDAR can be assessed by calculating the percentage of match between the MPCs from the SECL measurements at each stage of correction and the ground truth.

In this study, MPC matching is treated as an assignment problem and solved using the Hungarian algorithm~\cite{kuhn1955,NISTpaper}. The Hungarian algorithm aims to find the best one-to-one correspondence between all the elements of any two sets by minimizing the distance between all possible pairings collectively and iteratively. The sum of the distances between each pairing after an iteration is the total matching cost at that iteration, and after the matching process is completed, the total cost attains its minimum. In this subsection, we describe the Hungarian algorithm based MPC matching, and the numerical results from measurements will be presented in the next section. 

The Hungarian algorithm based MPC matching method has the following inputs: MPCs extracted from the target and source measurements, and the definition of the cost of assigning a source MPC to a target MPC. Here, the target measurement refers to the ground truth.

MPCs can be represented by a six-dimensional vector containing the AoA and the AoD azimuth and elevation angles, the delay, and the path gain as
\begin{equation}
x_i=({\theta}^{\mathrm{AoD,Az}}_i, {\theta}^{\mathrm{AoD,El}}_i, {\theta}^{\mathrm{AoA,Az}}_i, {\theta}^{\mathrm{AoA,El}}_i, \tau_i, \alpha_i),
\end{equation}
for $i=1,\dots,N_\mathrm{t}$ and $i=1,\dots,N_\mathrm{s}$, where $N_\mathrm{t}$ and $N_\mathrm{s}$ are the total number of MPCs in the target and source domains, respectively. The distance between the $l$th dimensions of the vectors representing the target MPC $i$ and the source MPC $j$ is defined as 
\begin{equation}
\Delta^l_{i,j}=\mid x^{l}_{\mathrm{t},i}-x^{l}_{\mathrm{s},j}\mid.
\end{equation}

To account for the unit and range differences across the dimensions, the distance along each dimension is normalized as
\begin{equation}
\hat{\Delta}_{i,j}^l=\frac{\Delta^l_{i,j}-\min{\bm{\Delta}^l_{j}}}{\max{\bm{\Delta}^l_{j}}-\min{\bm{\Delta}^l_{j}}},
\end{equation}
where ${\bm{\Delta}^l_{j}}=\{\Delta^l_{i,j}:1\le i \le N_\mathrm{t}\}$, for $j=1,\dots,N_\mathrm{s}$ and $l=1,\dots,6$. Then the cost of assigning the $i$th target MPC to the $j$th source MPC is calculated by 
\begin{equation}
\label{eq:cost}
c_{i,j}=\|\hat{\bm{\Delta}}_{i,j}\|,
\end{equation}
where ${\hat{\bm{\Delta}}_{i,j}}=\{\hat{{\Delta}}^l_{i,j}:1\le l \le 6\}$, and $\|\cdot\|$ is the $\ell_2$-norm operator. 
Having calculated the cost values, the assignment (or the MPC matching) problem can be expressed as
\begin{subequations}
\label{eq:Matching}
\begin{flalign}
\min &\sum_{i=1}^{N_\mathrm{t}}\sum_{j=1}^{N_\mathrm{s}}c_{i,j}\gamma_{i,j} & \\
\textrm{s.t.}&  \sum_{i=1}^{N_\mathrm{t}}\gamma_{i,j} = 1, \quad \forall j=1,\dots,N_\mathrm{s}, \quad \text{if}\ N_\mathrm{t} \leq N_\mathrm{s} \\
& \sum_{j=1}^{N_\mathrm{s}}\gamma_{i,j} = 1, \quad \forall i=1,\dots,N_\mathrm{t}, \quad \text{if}\   N_\mathrm{t} > N_\mathrm{s} \\
& \gamma_{i,j} \in \{0,1\}
\end{flalign}
\end{subequations}
for which $\gamma_{i,j}$ is equal to 1 if there is a match between the target MPC $i$ and the source MPC $j$, and 0 otherwise. The constraints in~(\ref{eq:Matching}) assure that an MPC from one domain can be matched with only a single MPC from the other domain, and the total number of matching MPCs is limited and equal to $\min(N_\mathrm{t},N_\mathrm{s})$. The effectiveness of the proposed delay correction methods will be validated with the numerical results of the Hungarian algorithm based MPC matching.

\section{Measurement Results}
\label{Sec:Results}
To see the effect of the proposed CCD and CDEDAR, two metrics are used: percentage of the matched MPCs extracted from the target and the source measurements, and the total power of the matched MPCs in the corresponding domains. Measurements are carried out in the lab environment shown in Fig.~\ref{setup:fig} for both cases, where a single clock is used for the TX and the RX, and then the TX and the RX clocks are disconnected. To make a fair comparison between the target and source measurements, the environment was kept stationary, i.e., there were no people in the lab, and the channel sounder is controlled through remote access.
\subsection{Quality of Matches}
Original Hungarian algorithm only aims to solve the assignment problem with the possible minimum cost (see the goal function in~(\ref{eq:Matching}a)). Therefore, it is possible that the matches are not as close as desired. To solve this problem, some criterion should be defined and checked before finding the matches.

\begin{figure*}[t]
	\begin{subfigure}{0.46\textwidth}
	\centering
	\includegraphics[width=\textwidth]{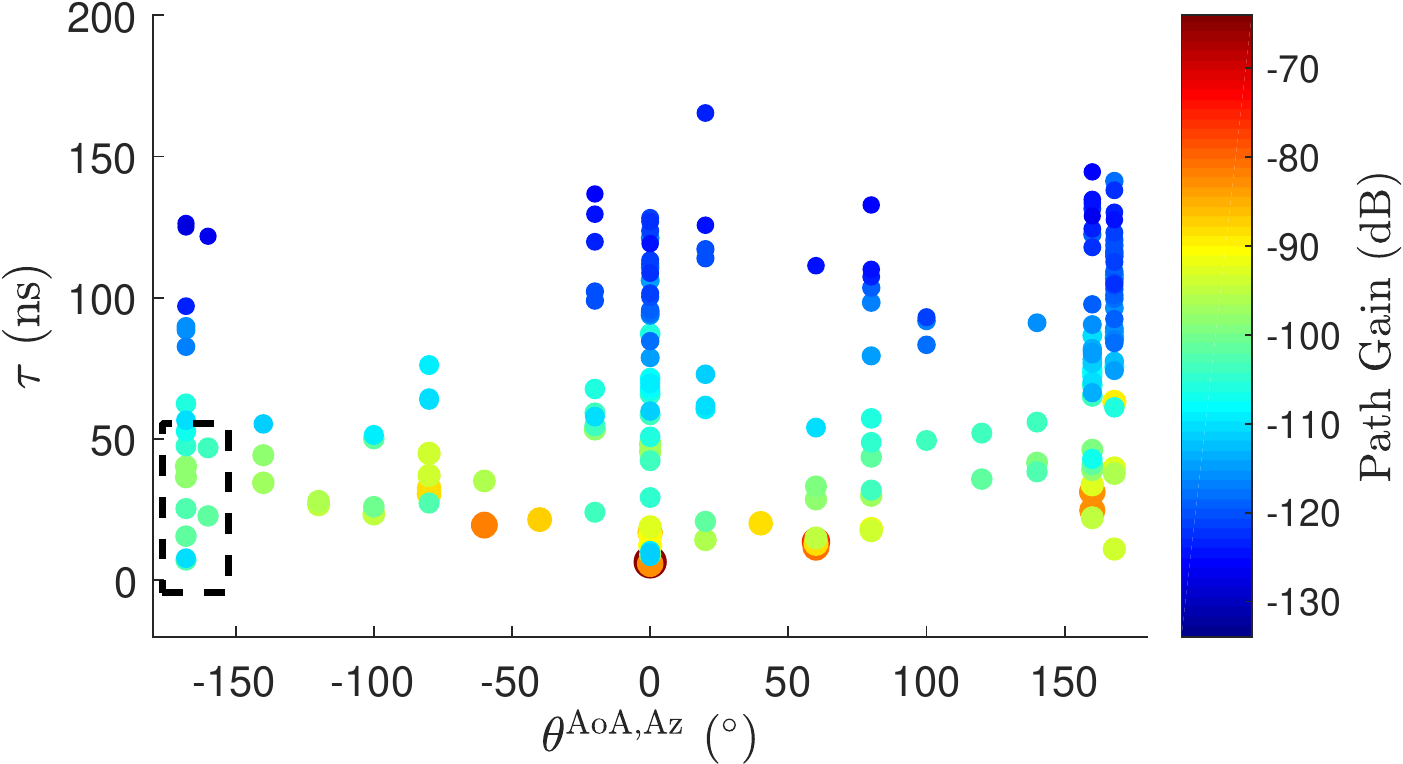}
	\caption{SICL After CDEDAR}
    \end{subfigure}			
	\begin{subfigure}{0.46\textwidth}
	\centering
    \includegraphics[width=\textwidth]{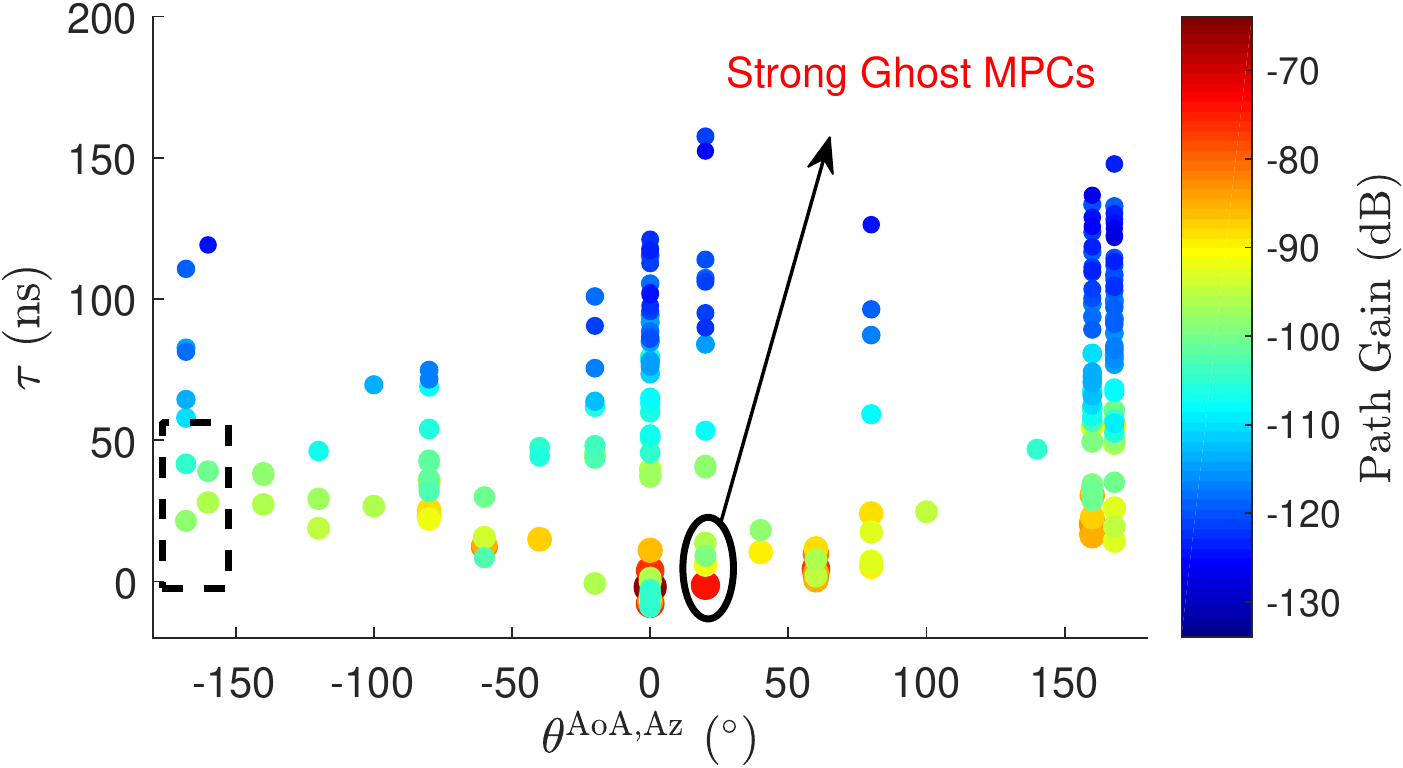}
	 \caption{SECL Raw}
     \end{subfigure}
     \begin{subfigure}{0.46\textwidth}
	\centering
	\includegraphics[width=\textwidth]{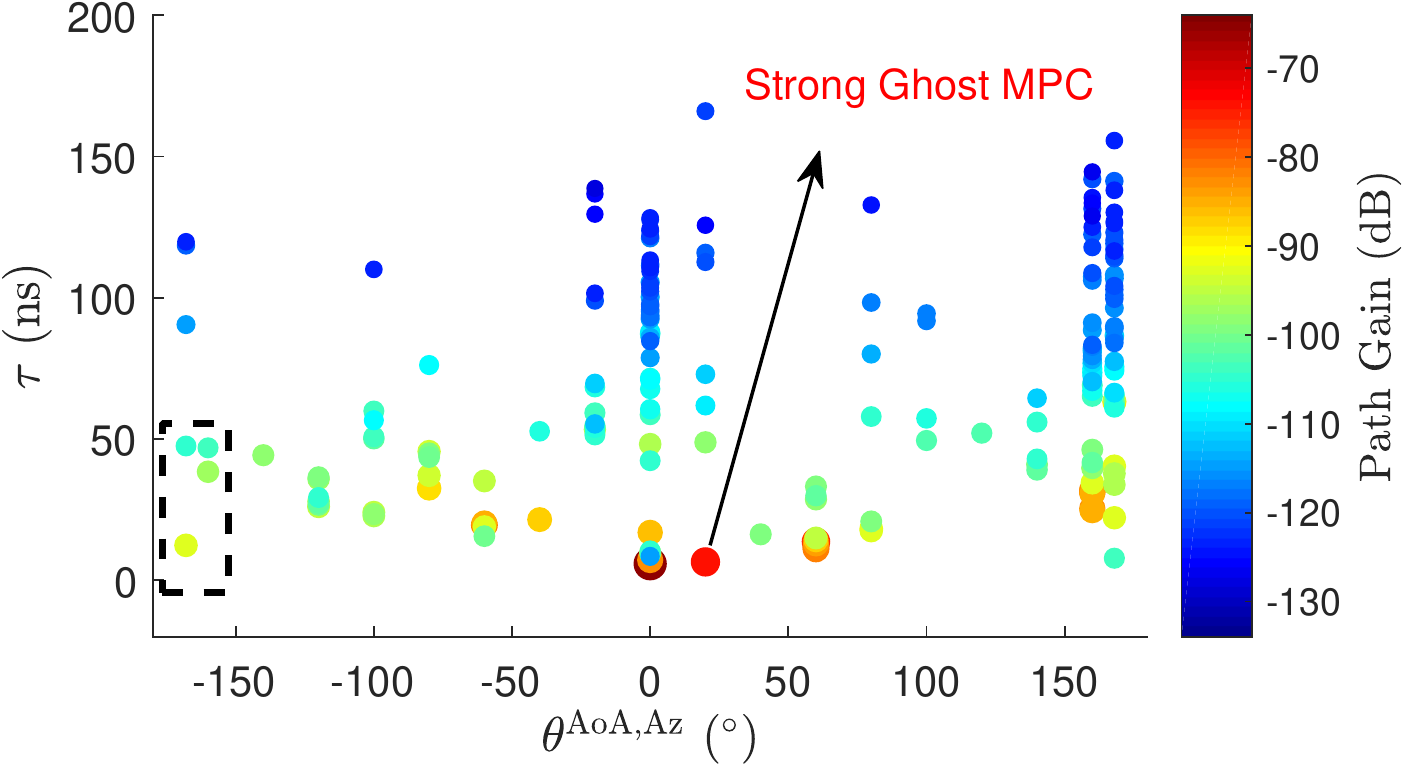}
	\caption{SECL After CCD}
    \end{subfigure}			
    \hspace{15mm}
	\begin{subfigure}{0.46\textwidth}
	\centering
    \includegraphics[width=\textwidth]{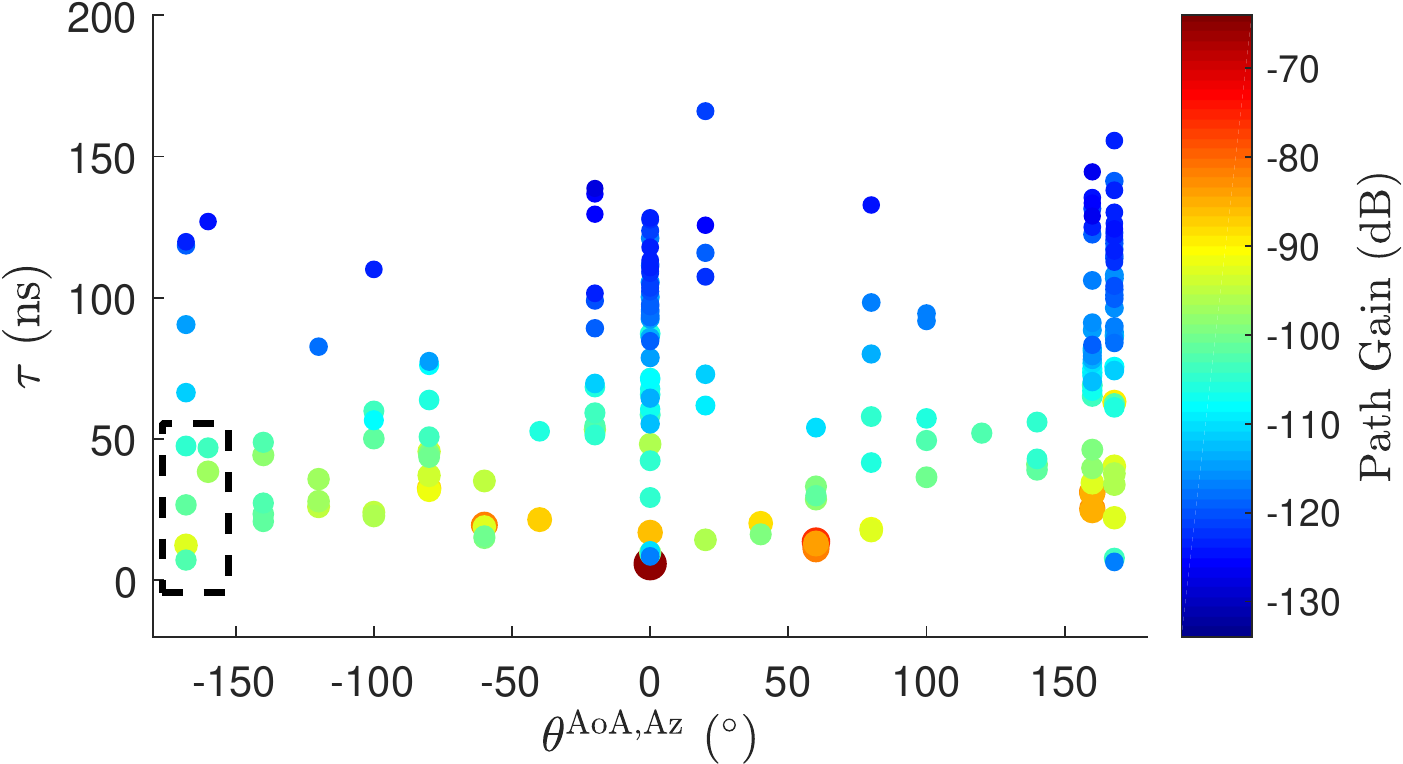}
	 \caption{SECL After CCD \& CDEDAR}
     \end{subfigure}
     \caption{Power in the delay-AoA azimuth plane (departure angles omitted) for the (a) single-clock measurement with antenna rotation errors corrected, (b) separate-clock raw measurement, (c) separate-clock measurement after clock drift corrected, and (d) separate-clock measurement after both errors corrected. Greater power values are represented by larger circles.}
     \label{Fig:DelayPowerAoAHor}
      \vspace{-1mm}
\end{figure*}

\begin{figure*}[h!]
	\begin{subfigure}{0.46\textwidth}
	\centering
	\includegraphics[width=\textwidth]{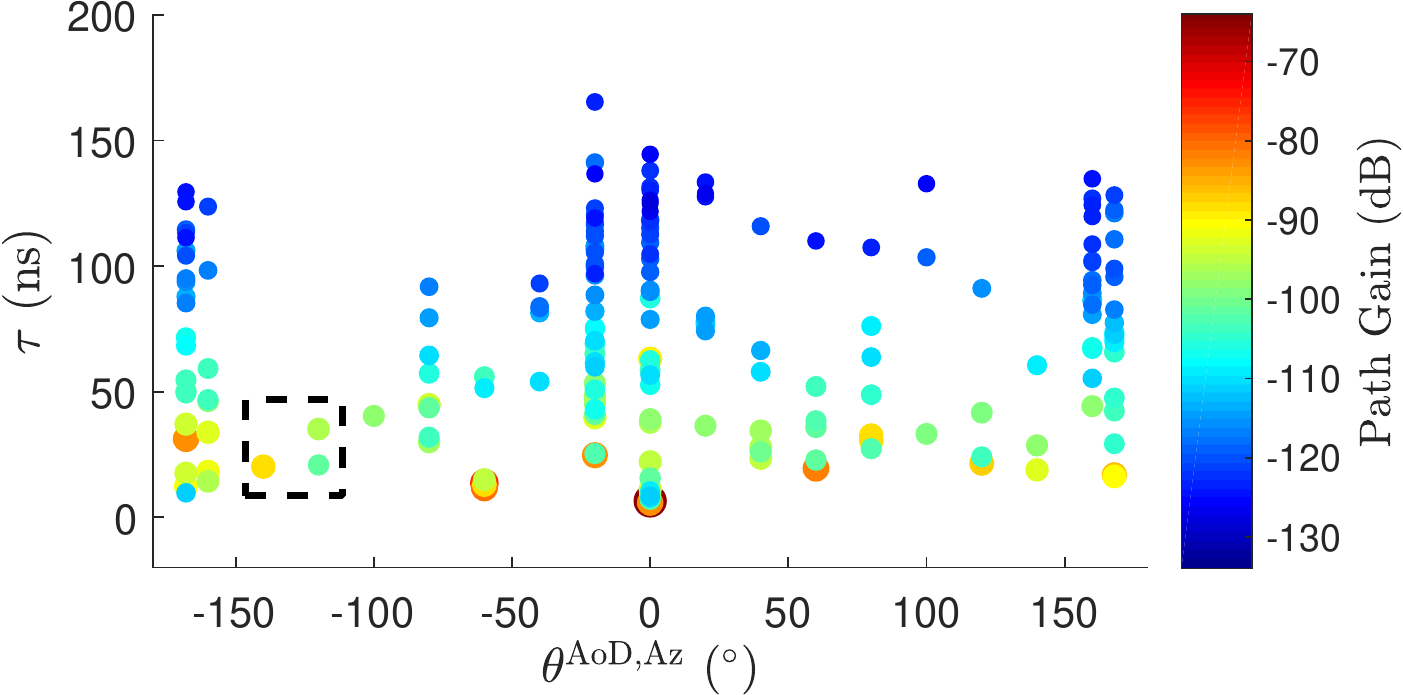}
	\caption{SICL After CDEDAR}
    \end{subfigure}			
	\begin{subfigure}{0.46\textwidth}
	\centering
    \includegraphics[width=\textwidth]{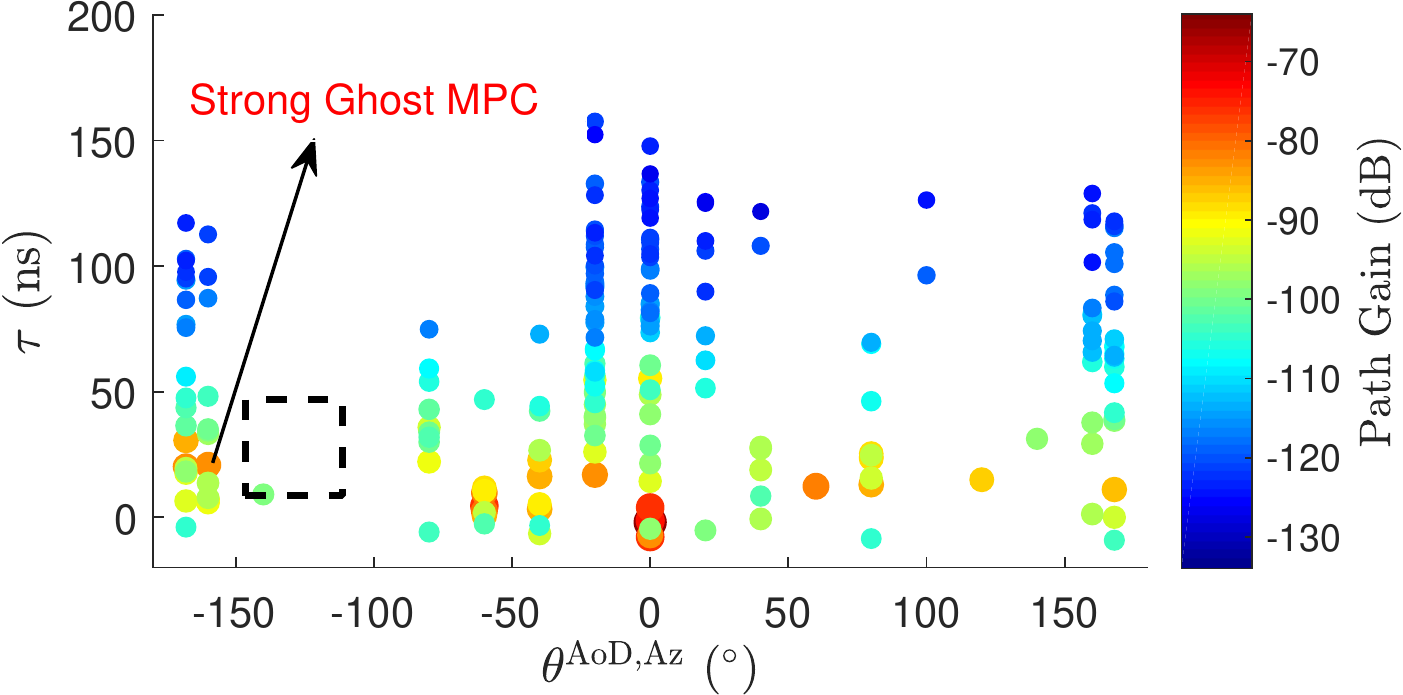}
	 \caption{SECL Raw}
     \end{subfigure}
     \begin{subfigure}{0.46\textwidth}
	\centering
	\includegraphics[width=\textwidth]{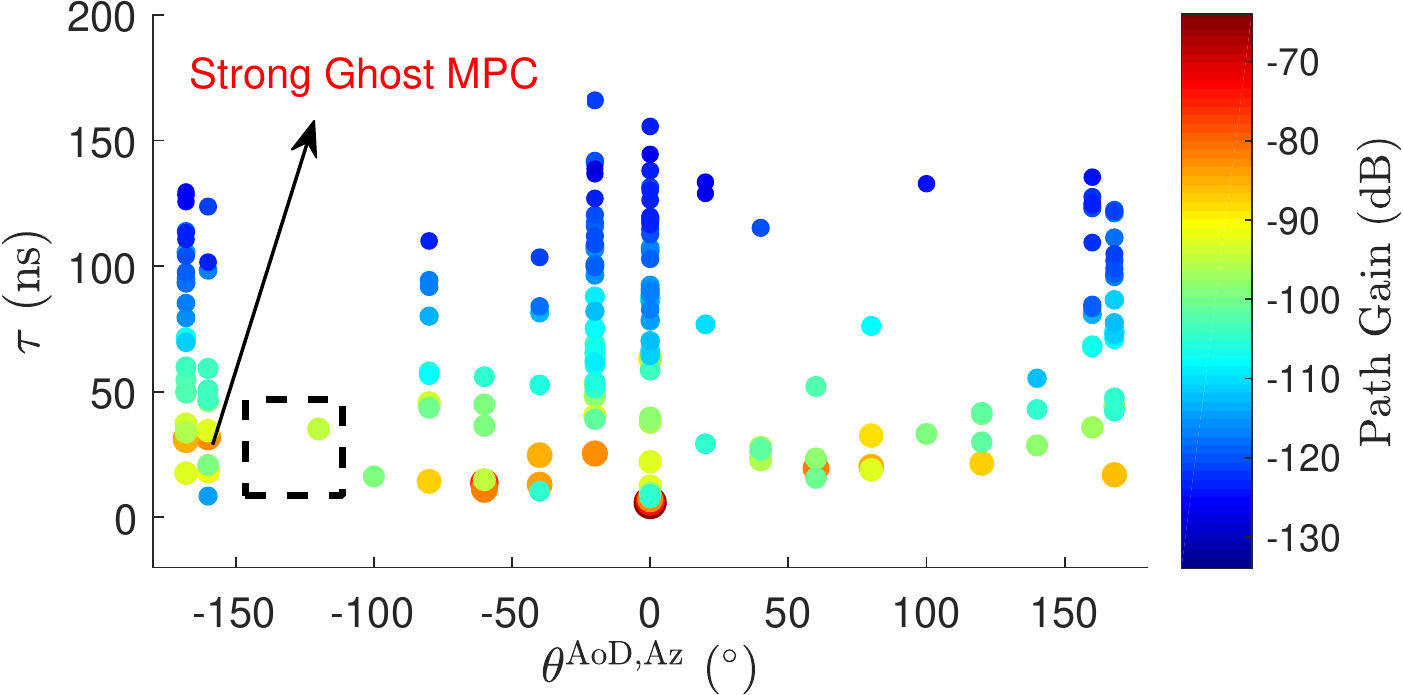}
	\caption{SECL After CCD}
    \end{subfigure}		
    \hspace{15mm}
	\begin{subfigure}{0.455\textwidth}
	\centering
    \includegraphics[width=\textwidth]{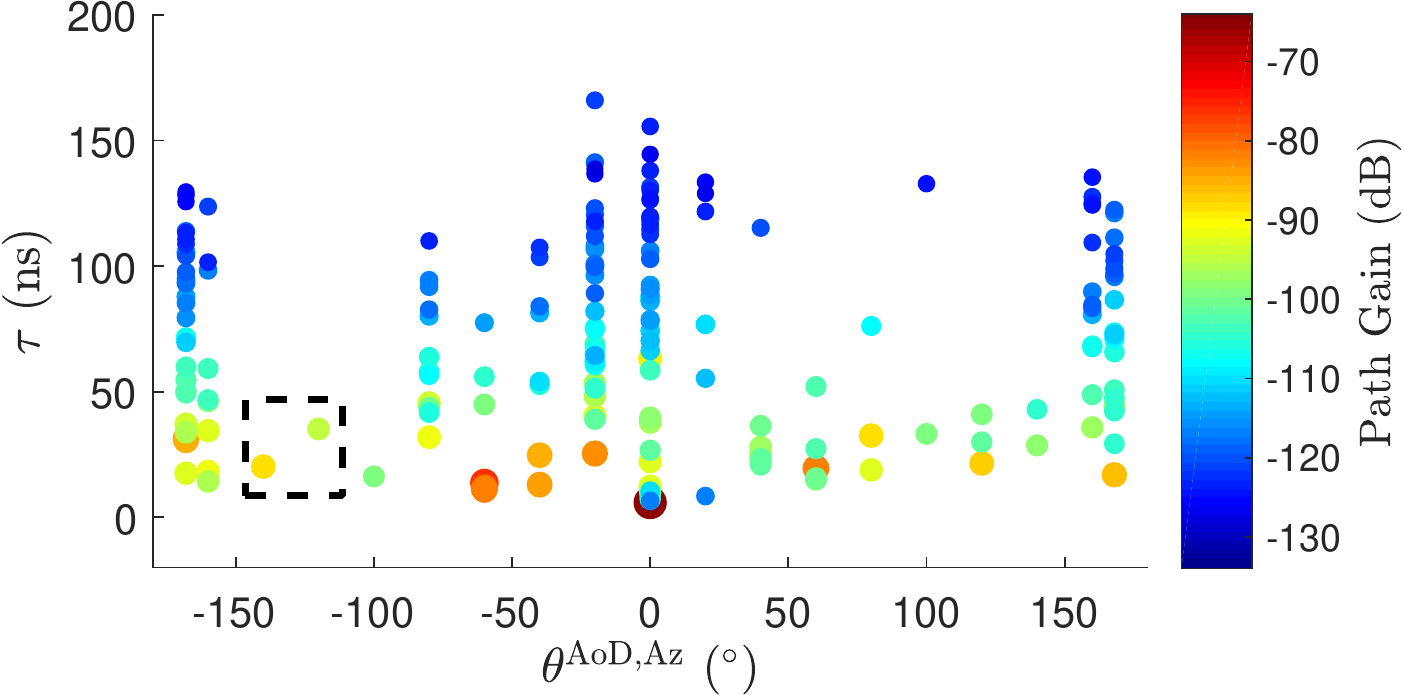}
	 \caption{SECL After CCD \& CDEDAR}
     \end{subfigure}
     \caption{Power in the delay-AoD azimuth plane (arrival angles omitted) for the (a) single-clock measurement after antenna rotation errors corrected, (b) separate-clock raw measurement, (c) separate-clock measurement after clock drift corrected, and (d) separate-clock measurement after both errors corrected. Greater power values are represented by larger circles.}
     \label{Fig:DelayPowerAoDHor}
\end{figure*}

In this study, a match between the target MPC $i$ and the source MPC $j$ is considered as \textit{valid} if all the following conditions are met: 
\begin{enumerate}
    \item  $\Delta^l_{i,j}  < 22.5^{\circ}$, for $l=1,\dots,4$,
    \item $\Delta^5_{i,j}  < 5$ samples ($=3.255$~ns), and
    \item $\Delta^6_{i,j}  < 4$ dB.
\end{enumerate}
The first condition defines the maximum difference allowed in all the angular dimensions of the two MPCs, whereas the second and third conditions specify the error tolerance in the delay and path gain dimensions, respectively. If at least one of the above conditions is not met for a target and source MPC pair, then the matching cost is set to $-\infty$ for these two MPCs so that they are never matched with each other. Besides, if the cost of all the possible matches for a certain MPC in one of the domains turns out to be equal to $-\infty$, then that MPC is labeled as unassigned and discarded from consideration in the matching process.

Since the resolution is higher in the delay and path gain dimensions of the MPC space compared to the angular dimensions (due to the setup used), the delay and the path gain can be considered as more reliable parameters in defining an MPC. Therefore, these two parameters should have more weight in determining the matches. To account for that, the cost definition in~(\ref{eq:cost}) is updated as 
\begin{equation}
c_{i.j} =\sqrt{\sum_{l=1}^{6} w_l \left(\hat{\Delta}_{i,j}^{l}\right)^2},
\end{equation}
where $w_l$ is the weight of the value (in the total cost) in dimension $l$, and it is set to $1/4$ for $l=1,\dots,4$ and 1 for both $l=5$ and $l=6$.



\begin{table}[t]
\footnotesize
\renewcommand{\arraystretch}{1.2}
\caption {MPC Matching Results for the Source Measurements with and without Correction(s).}
\label{Tab:MPCmatching1}
\resizebox{0.49\textwidth}{!}
{\begin{tabular}{m{2.1cm}ccccc}
\hline
\multicolumn{5}{l}{Total number of target (SICL+CDEDAR) MPCs: 201}\\\hline
 & & & & \multicolumn{2}{c}{Matched power}\\
\multirow{2}{*}{Source}&$\#$ extracted & Matched & Missed& \multicolumn{2}{c}{(\%)}\\\cline{5-6} 
&MPCs & MPCs (\%)& MPCs (\%) & Target & Source \\\hline
SECL & 215   &28.36   & 71.64 & 29.95&29.10 \\
SECL+CCD   & 204  & 70.65   & 29.35 & 72.24&71.03\\
SECL+CCD\&CDEDAR  &203   & 74.13   & 25.87 & 74.24&73.55\\\hline
\end{tabular}}
\end{table}

\begin{table*}[t!]
\footnotesize
\renewcommand{\arraystretch}{1.2}
\caption {Parameters of the the Top 15 Strongest MPCs of the Target and the Source Measurements and Corresponding Matching Results.}
\label{Tab:MPC15}
\resizebox{\textwidth}{!}{
\begin{tabular}{c|clc|cclc|cclc}
\hline
\multicolumn{1}{l|}{}    & \multicolumn{3}{c|}{Single-clock measurement after CDEDAR} & \multicolumn{4}{c|}{Separate-clock measurement after CCD}& \multicolumn{4}{c}{Separate-clock measurement after CCD\&CDEDAR}\\ \hline
\multicolumn{1}{c|}{} & ${\theta}^{\mathrm{AoD,Az}}_i$ / ${\theta}^{\mathrm{AoD,El}}_i$ /  & & \multicolumn{1}{l|}{} & & ${\theta}^{\mathrm{AoD,Az}}_i$ / ${\theta}^{\mathrm{AoD,El}}_i$ /  & & \multicolumn{1}{l|}{} & & ${\theta}^{\mathrm{AoD,Az}}_i$ / ${\theta}^{\mathrm{AoD,El}}_i$ /  & & \multicolumn{1}{l}{} \\ 
\multicolumn{1}{c|}{MPC} & ${\theta}^{\mathrm{AoA,Az}}_i$ / ${\theta}^{\mathrm{AoA,El}}_i$ & \multicolumn{1}{c}{$\tau_i$} & \multicolumn{1}{c|}{$\alpha_i$} & Match &  ${\theta}^{\mathrm{AoA,Az}}_i$ / ${\theta}^{\mathrm{AoA,El}}_i$ & \multicolumn{1}{c}{$\tau_i$} & \multicolumn{1}{c|}{$\alpha_i$} & Match &  ${\theta}^{\mathrm{AoA,Az}}_i$ / ${\theta}^{\mathrm{AoA,El}}_i$  & \multicolumn{1}{c}{$\tau_i$} & \multicolumn{1}{c}{$\alpha_i$} \\ 
\multicolumn{1}{c|}{\# (i)} & ($^\circ$) & (ns) & \multicolumn{1}{c|}{(dB)} & (i) &  ($^\circ$) & (ns) & \multicolumn{1}{c|}{(dB)} & (i) &  ($^\circ$)  & (ns) & \multicolumn{1}{c}{(dB)} \\ 
\hline
1         & 0 / 0  / 0       / 0        & 10         & -66.01     &(1)  & 0         / 0         / 0         / 0         & 9          & -66.01     &(1)  & 0         / 0         / 0         / 0         & 9          & -66.01     \\2            & -60  / 0                 / 60      / 0                                                   & 21         & -77.01     &\xmark  & 0         / 0         / 20        / 0         & 10         & -74.89     &(2)  & -60       / 0         / 60        / 0         & 21         & -76.89     \\
3                & -60 / 0               / 60 / -20            & 18         & -81.39     &(2)  & -60       / 0         / 60        / 0         & 21         & -76.89     &(3)  & -60       / 0         / 60        / -20       & 18         & -81.25     \\
4& 60 / 0 / -60   / 0                 & 30         & -81.67     &(3)  & -60       / 0         / 60        / -20       & 18         & -81.25     &(4)  & 60        / 0         / -60       / 0         & 30         & -81.64     \\
5                & 0                   / -20 / 0 / -20 & 9          & -82.74     &(4) & 60        / 0         / -60       / 0         & 30         & -81.64     &\xmark  & -60       / -20       / 60        / 0         & 17         & -82.46     \\
6& -168               / 0   / 160/ 0                & 48         & -82.79     &\xmark  & -60       / -20       / 60        / 0         & 17         & -82.46     &(6)  & -168   / 0         / 160       / 0         & 48         & -82.63     \\
7 & -20 / 0/ 160/ 0  & 38         & -83.46     &(6)  & -168   / 0         / 160       / 0         & 48         & -82.63     &(7)  & -20       / 0         / 160       / 0         & 39         & -83.28     \\
8                                                    & 168                                              / 0                                                    / 0                                                   / 0                                                   & 26         & -86.22     &\xmark  & -160      / 0         / 160       / 0         & 49         & -83.11     &(10)  & -40       / 0         / 60        / 0         & 20         & -84.31     \\
9                                                    & 120                                                 / 0                                                    / -40                                                 / 0                                                   & 33         & -87.81     &(7)  & -20       / 0         / 160       / 0         & 39         & -83.28     &\xmark  & -40       / 0         / 160       / 0         & 38         & -84.81     \\
10                                                   & -60                                                 / -20                                                  / 60                                                  / -20                                                 & 20         & -88.13     &\xmark  & 0         / 20        / 0         / 20        & 11         & -83.51     &\xmark  & -168   / -20       / 160       / 0         & 47         & -85.53     \\
11                                                   & 80                       / 0                 / -80                         / 0                      & 47         & -88.14     &\xmark  & -40       / 0         / 60        / 0         & 20         & -84.31     &(8) & 168    / 0         / 0         / 0         & 26         & -86.02     \\
12       & -140      / 0         / 40  / 0       & 31  & -88.92     &\xmark  & -40       / 0     / 160       / 0         & 38    & -84.81     &(9) & 120       / 0     / -40       / 0         & 33     & -87.84     \\
13 & 80    / 0                                / -80     / 0    & 50         & -89.08     &\xmark  & 80        / 0         / -60       / 0         & 31         & -84.89     &(12)  & -140      / 0         / 40        / 0         & 31         & -88.64     \\
14   & 0  / 0                              / 168    / 0     & 97         & -90.22     &\xmark  & -168   / -20       / 160       / 0         & 47         & -85.53     &(13)  & 80        / 0         / -80       / 0         & 50         & -88.88     \\
15       & -168         / -20          / 0    / 0     & 19      & -90.42     &(8)  & 168    / 0      / 0         / 0         & 26      & -86.02     &(14)  & 0         / 0       / 168    / 0         & 97      & -90.22     \\ \hline
\end{tabular}}
\end{table*}
\subsection{Performance of the Clock Drift and Antenna Error Correction}
The MPCs extracted after applying CDEDAR to SICL measurements and the MPCs extracted from the SECL measurements before and after CCD (and CDEDAR) is (are) applied are shown in Fig.~\ref{Fig:DelayPowerAoAHor} and Fig.~\ref{Fig:DelayPowerAoDHor} in the delay-AoA azimuth and the delay-AoD azimuth planes, respectively. As an example of the contribution of the correction processes to the accuracy of the extracted MPCs, a group of MPCs are shown in dashed boxes in Fig.~\ref{Fig:DelayPowerAoAHor}(a)-(d) for the delay-AoA azimuth plane. The number of missed MPCs in the labeled regions decreases after each correction applied to the SECL measurements. We also see that some of the strong ghost MPCs are eliminated after applying CCD and CDEDAR. Similar observations can be made from Fig.~\ref{Fig:DelayPowerAoDHor} where we show the MPCs in the delay-AoD azimuth plane.

Matching results for the target and source MPCs are presented in Table~\ref{Tab:MPCmatching1}. 
Total number of the target MPCs is 201, and the closest result is obtained for the case when both errors are corrected in the SECL measurement, which is 203. It is observed that the percentage of the matched MPCs is increased from 28.36\% to 74.13\% after implementing the CCD and CDEDAR in the SECL measurement. Moreover, the ratio of the matched power is increased from 29.95\% and 29.10\% to 74.24\% and 73.55\% in the target and source domains, respectively, after the correction of both errors in the SECL measurement. 

\begin{figure}[t]
	\begin{subfigure}{0.24\textwidth}
	\centering
	\includegraphics[width=\textwidth]{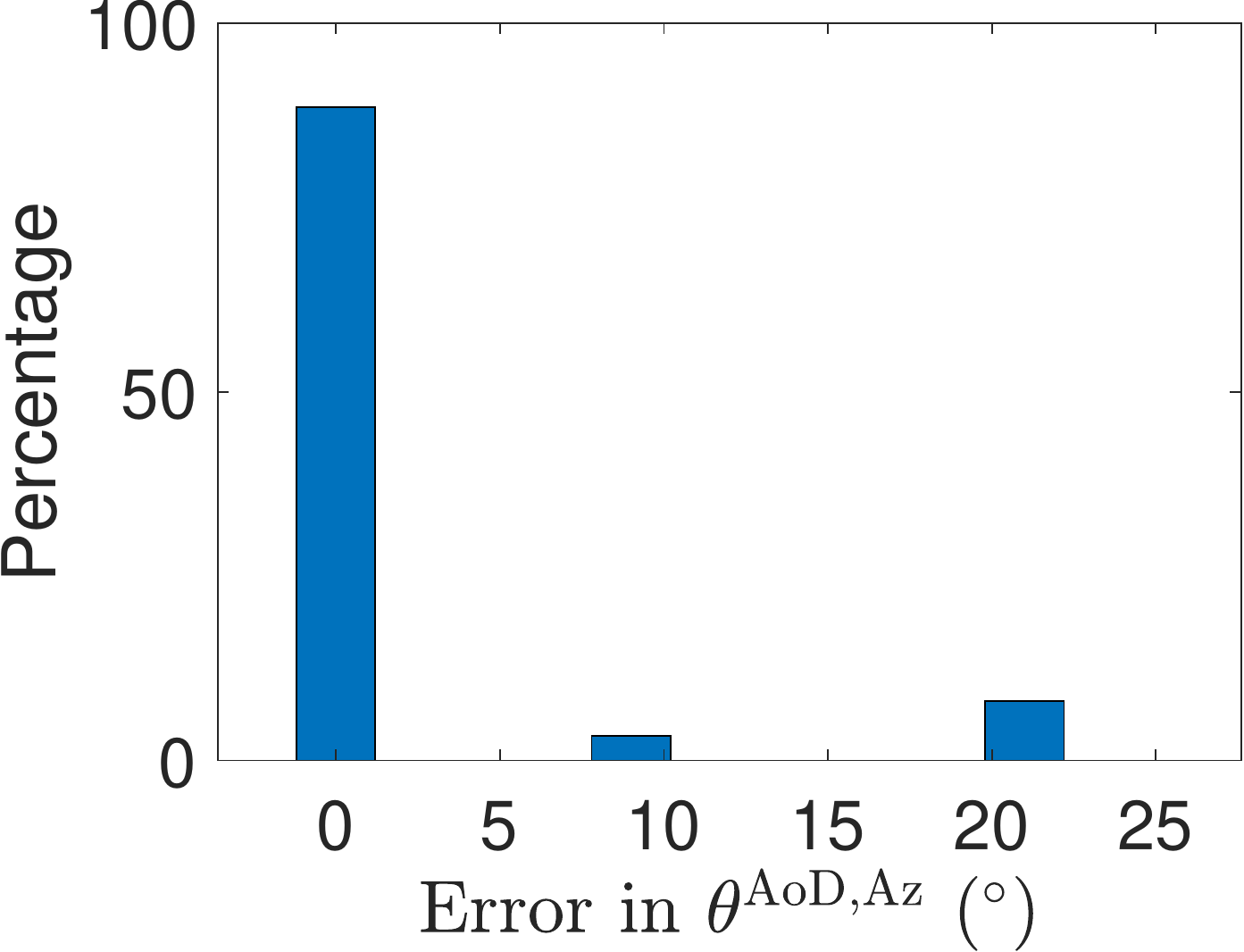} 
	\caption{AoD azimuth}
    \end{subfigure}			
	\begin{subfigure}{0.24\textwidth}
	\centering
    \includegraphics[width=\textwidth]{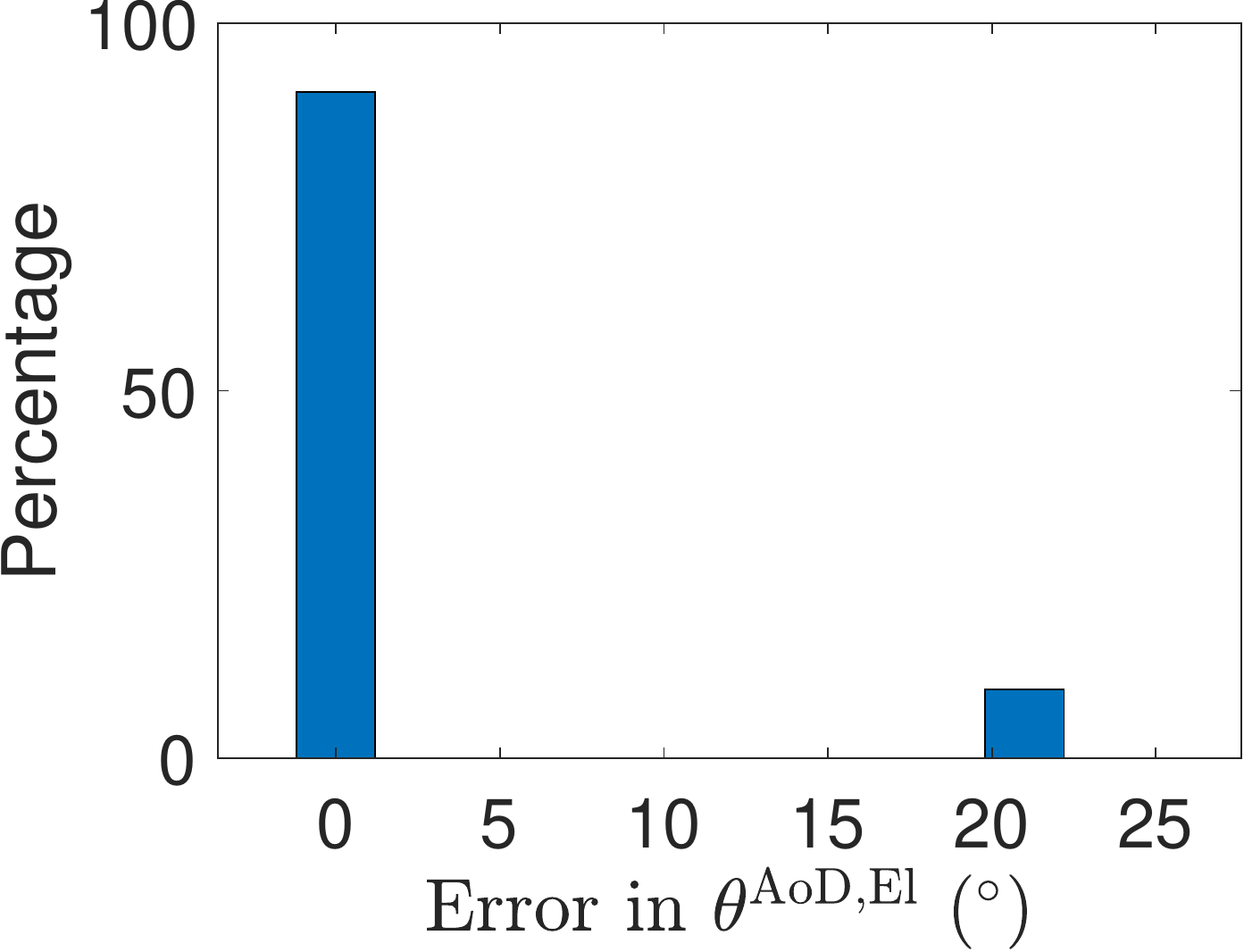}
	 \caption{AoD elevation}
     \end{subfigure}
     \begin{subfigure}{0.24\textwidth}
	\centering
	\includegraphics[width=\textwidth]{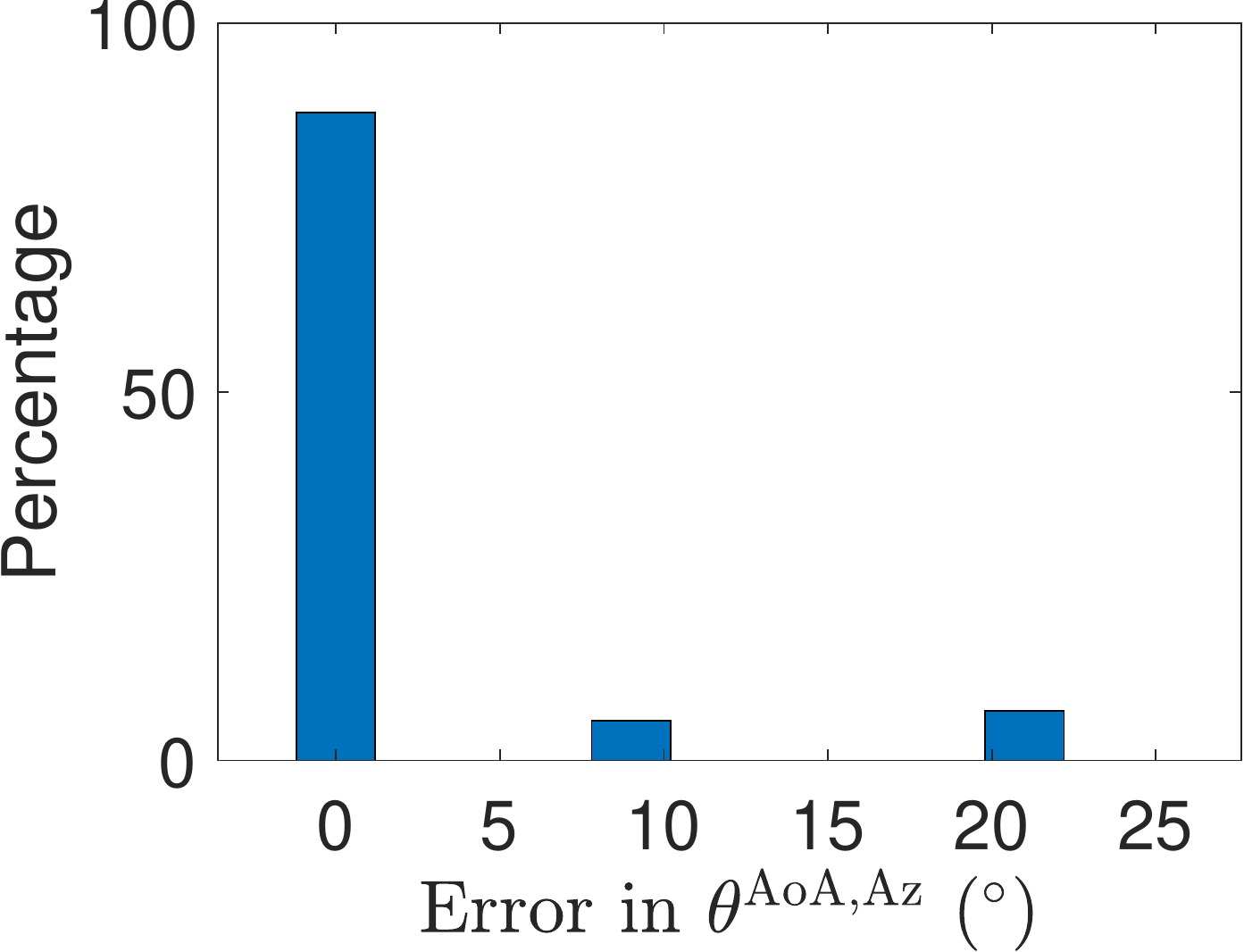} 
	\caption{AoA azimuth}
    \end{subfigure}			
	\begin{subfigure}{0.24\textwidth}
	\centering
    \includegraphics[width=\textwidth]{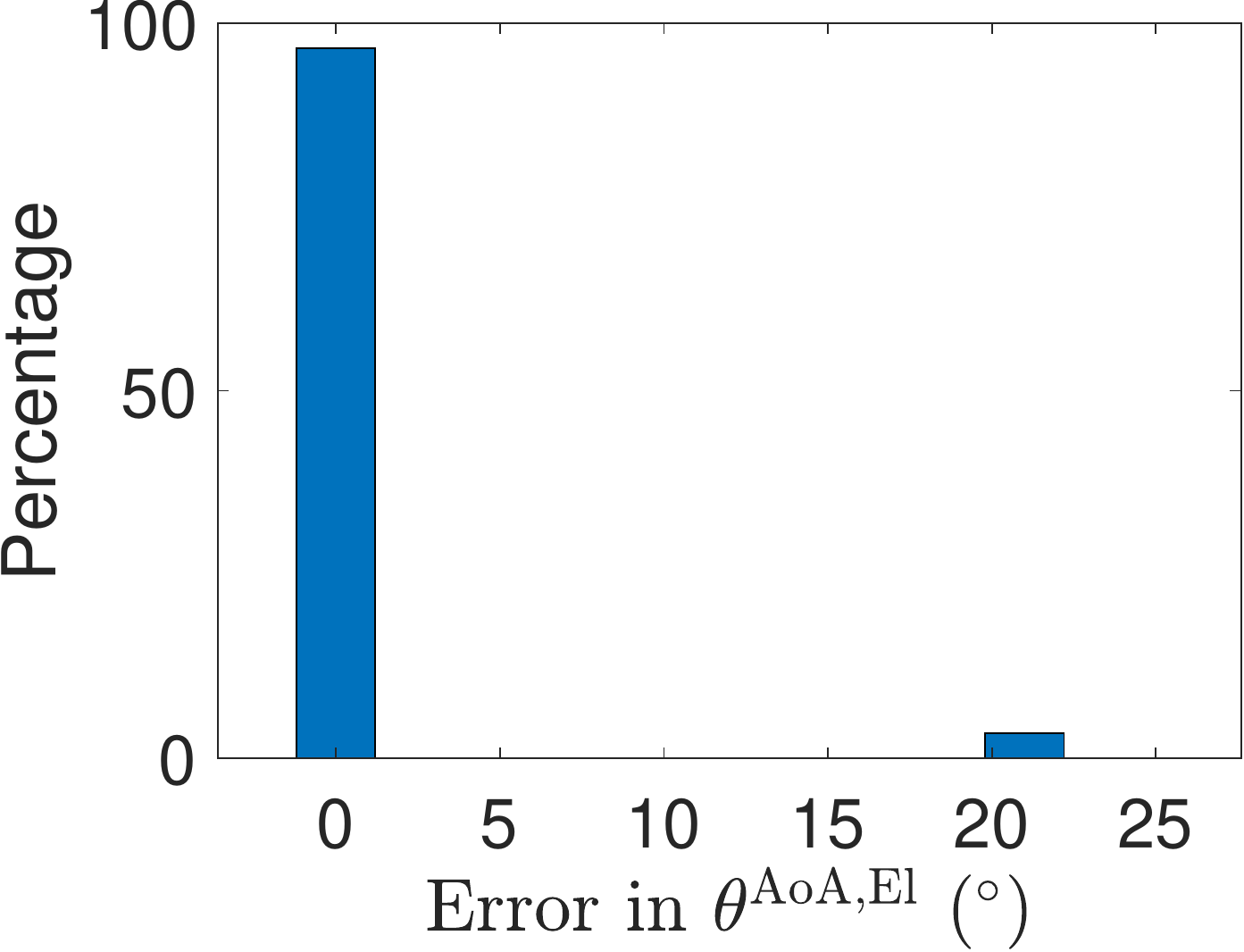}
	 \caption{AoA elevation}
     \end{subfigure}
     \begin{subfigure}{0.24\textwidth}
	\centering
	\includegraphics[width=\textwidth]{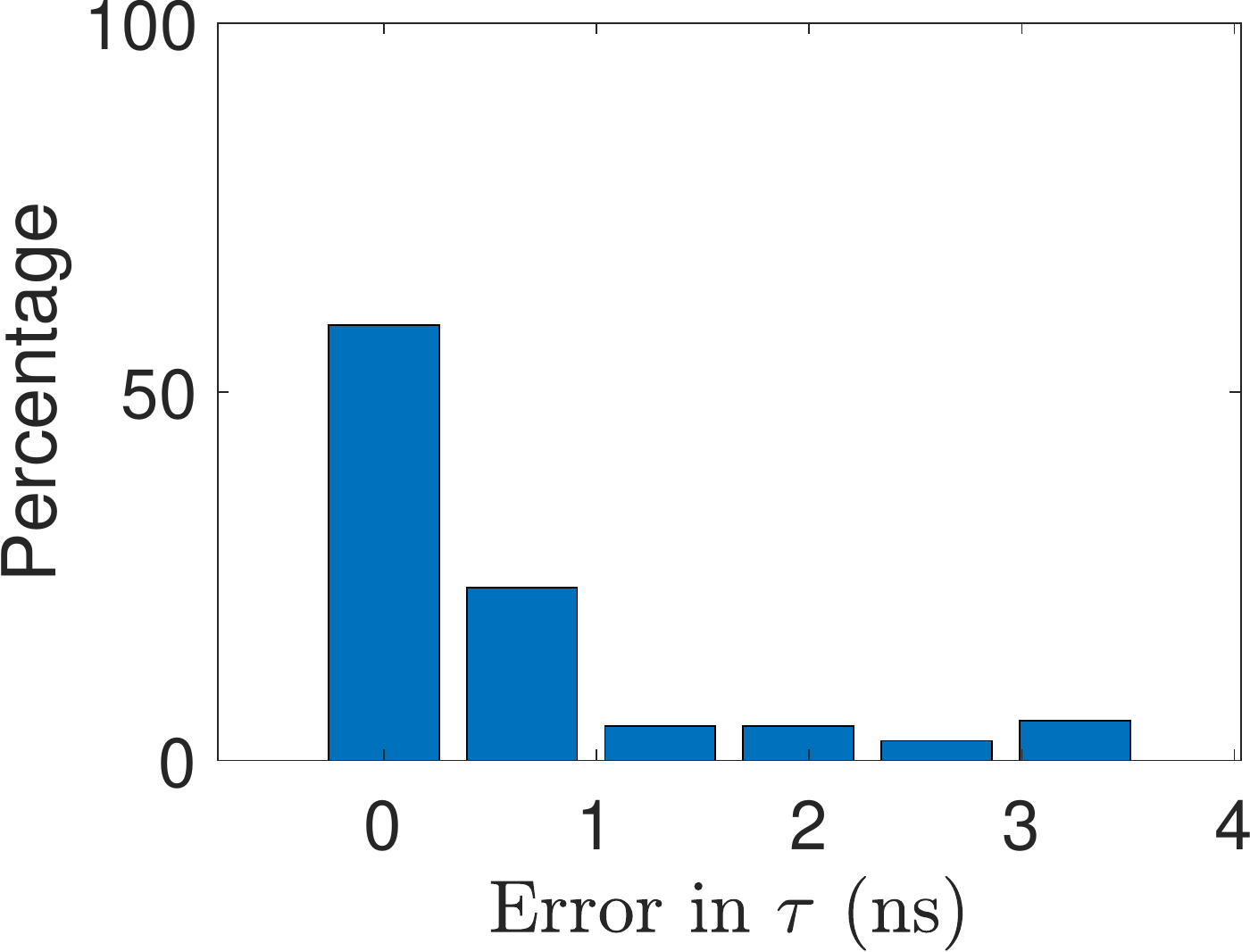} 
	\caption{Delay}
    \end{subfigure}			
	\begin{subfigure}{0.24\textwidth}
	\centering
    \includegraphics[width=\textwidth]{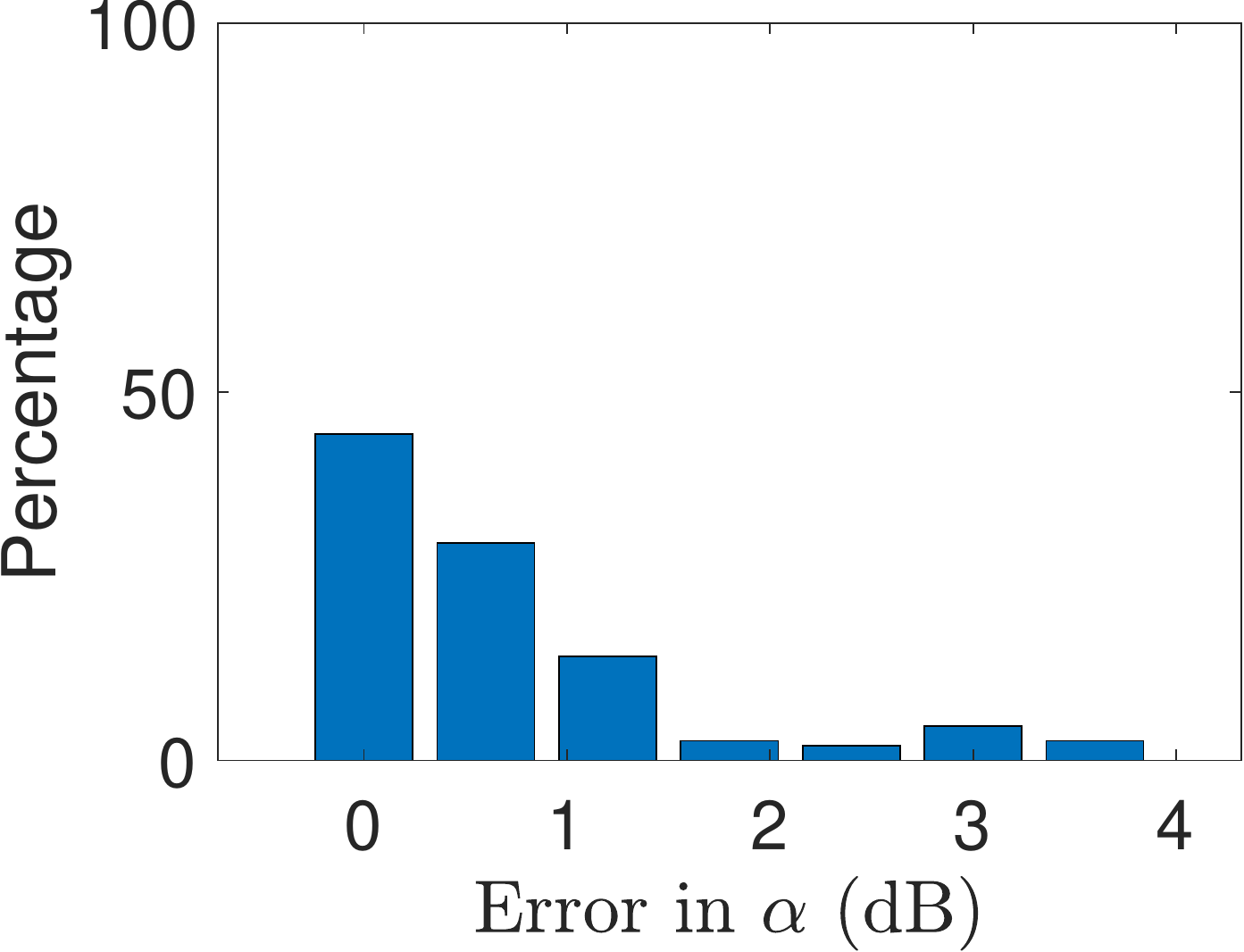}
	 \caption{Path gain}
     \end{subfigure}
     \caption{Percentage histogram of errors between the parameters of the matched MPCs of the target and corrected source measurements. }\label{Fig:ErrorHist}
\end{figure}

The percentage histograms of errors in each dimension between the MPCs of the target and corrected source measurements are shown in Fig.~\ref{Fig:ErrorHist}. It can be observed that almost all the MPCs that are matched have the same angular parameters and differ less than 1 ns in the delay and 2 dB in the path gain. \looseness=-1

Finally, the parameters of the top 15 strongest MPCs of the target and source measurements are presented in Table~\ref{Tab:MPC15}. If a match is found for an MPC from the source domain, then it is indicated by the index of the corresponding MPC from the target domain; otherwise, it is indicated using the cross mark.  Only 7 of the 15 MPCs in the SECL measurement after CCD are matched with a target MPC whereas the number of matches is 12 in the case of the SECL measurement after CCD and CDEDAR.

\section{Conclusions}
\label{sec:Conc}
In this paper, we introduced our modified NI-based 28~GHz channel sounder to measure the PADP of the mmWave propagation channel. We described an MPC extraction method, and CDC and CDEDAR algorithms to correct the errors while extracting the MPCs from measurements. The results show that the MPC matching between the SICL measurement and the SECL measurements increases from 28.36\% in SICL raw measurement to 74.13\% in SICL measurement after applying both CDC and CDEDAR. Therefore, for the accurate characterization of the channel, we conclude that the proposed correction algorithms should be applied to the measurements obtained from similar setups to ours. In the Hungarian based MPC matching method, we take a set of constraints into considerations to account for the quality of matches. This method can be used to compare the MPC parameters from different channel sounders and MPC extraction techniques for accuracy assessment. Our future work includes using more advanced correction algorithms, carrying out measurements in more complicated environments such as libraries and airports, and perform channel modeling at different mmWave bands.

\section*{Acknowledgment}
The authors would like to thank Camillo Gentile and Jack Chuang for their feedback on an earlier version of the paper and their help in developing the Hungarian algorithm based MPC matching method and Jeanne Quimby for her help in verification of our sounder. 
\balance
\bibliography{IEEEabrv,references}
\bibliographystyle{IEEEtran}
\end{document}